\documentclass[12pt]{article}

\usepackage[utf8]{inputenc}

\usepackage{amsmath,amssymb}
\usepackage{hyperref}
\usepackage{graphicx}
\usepackage{doi}
\usepackage{geometry}
\usepackage{setspace}
\usepackage{enumitem}
\usepackage{newunicodechar}
\newunicodechar{≠}{\neq}

\title{A Formal Rebuttal of “The Blockchain Trilemma: A Formal Proof of the Inherent Trade-Offs Among Decentralization, Security, and Scalability”}
\author{\Large Dr. Craig S. Wright \\
\small Faculty of Business, University of Exeter \\
\small \texttt{cw881@exeter.ac.uk}}
\date{}

\begin{document}

\maketitle

\begin{abstract}
\noindent This paper presents a comprehensive refutation of the so-called “blockchain trilemma,” a widely cited but formally ungrounded claim asserting an inherent trade-off between decentralisation, security, and scalability in blockchain protocols. Through formal analysis, empirical evidence, and detailed critique of both methodology and terminology, we demonstrate that the trilemma rests on semantic equivocation, misuse of distributed systems theory, and a failure to define operational metrics. Particular focus is placed on the conflation of topological network analogies with protocol-level architecture, the mischaracterisation of Bitcoin’s design—including the role of miners, SPV clients, and header-based verification—and the failure to ground claims in complexity-theoretic or adversarial models. By reconstructing Bitcoin as a deterministic, stateless distribution protocol governed by evidentiary trust, we show that scalability is not a trade-off but an engineering outcome. The paper concludes by identifying systemic issues in academic discourse and peer review that have allowed such fallacies to persist, and offers formal criteria for evaluating future claims in blockchain research. \footnote{A response to Mssassi, Souhail, and Anas Abou El Kalam. "The Blockchain Trilemma: A Formal Proof of the Inherent Trade-Offs Among Decentralization, Security, and Scalability." Applied Sciences 15, no. 1 (2024): 19. https://doi.org/10.3390/app15010019.}
\end{abstract}


\newpage

\section{Introduction}

The notion of a “blockchain trilemma” has become a recurring motif in both academic and popular discourse on distributed ledger technologies. It purports to articulate an inherent design constraint whereby no protocol can simultaneously maximise decentralisation, security, and scalability. This trilemma, often attributed to various reformulations but canonised through informal sources such as developer blogs and position papers, has been cited to justify architectural compromises across numerous systems. The claim is that one must sacrifice at least one of the three properties in the pursuit of the others — that this trade-off is not merely practical, but foundational and necessary.

Despite its prevalence, the trilemma lacks a formal derivation. It is not grounded in a coherent adversarial model, nor is it demonstrable through standard techniques in distributed systems theory, complexity analysis, or information security. Instead, it relies on equivocation of terms, circular reasoning, and selective misapplication of network analogies. Terms like “decentralisation” are used without reference to economic control or decision-making authority, while “security” is collapsed into participation redundancy. Scalability, meanwhile, is often mischaracterised as inherently incompatible with reduced verification or relay costs. In this paper, we contend that these conceptual failures are not peripheral, but central to the persistence of the trilemma as an unexamined dogma.

This document provides a comprehensive and formal critique of the trilemma framing, addressing both the semantic imprecision and the methodological flaws embedded in its use. We present a series of structural rebuttals, beginning with definitional clarification, moving through logical and empirical deconstructions, and culminating in a series of refutations grounded in formal properties of the Bitcoin protocol. The structure of the critique is cumulative: we begin by dismantling the claim’s reliance on semantic drift and network analogies, proceed to the improper use of formal language and the failure to define metrics or boundary conditions, and conclude with case analyses that demonstrate the empirical falsifiability of the trilemma in light of operational systems such as Bitcoin SV.

By re-establishing the formal, economic, and topological foundations of Bitcoin, this work reveals that the trilemma is not a theorem to be respected but a pseudoproblem to be discarded. Rather than a constraint on innovation, it serves as a rhetorical device that occludes the architectural and economic principles enabling scale, trust minimisation, and verification without redundancy. The goal herein is not merely to rebut a single paper or formulation but to reassert rigour in discussions of protocol design — to reject mystification in favour of testable structure and semantic clarity.

\subsection{Purpose and Scope of Rebuttal}

The purpose of this paper is to formally rebut the claim that Bitcoin — or blockchain-based systems more broadly — are subject to an inescapable trilemma, wherein decentralisation, scalability, and security cannot be simultaneously achieved. We reject this framing not merely on empirical grounds but through systematic deconstruction of its logical structure, semantic equivocations, and misapplied analogies. The trilemma is not a law, not a theorem, and not a constraint derivable from first principles. It is a narrative — one that survives through repetition rather than rigour.

The scope of this rebuttal is tripartite:

\begin{enumerate}
    \item \textbf{Formal Disassembly:} We analyse and disprove the claimed necessity of the trilemma by introducing proper definitions of decentralisation, scalability, and security. These terms are rendered in predicate logic and complexity bounds to replace their narrative and rhetorical usage with formally testable claims. Where proponents invoke informal metaphors and misleading analogies, we reinstate the primacy of formal model specification.

    \item \textbf{Architectural Correction:} We clarify the nature of Bitcoin as a deterministic protocol defined by rule-based execution and economic alignment. Decentralisation in Bitcoin is not a topological artefact but a consequence of protocol-enforced non-discretion. Security arises from the provability of hash-based evidence, not from trust in network actors. Scalability is an engineering outcome derived from bandwidth and latency constraints, not from mythical complexity limitations imposed by the number of participants.

    \item \textbf{Empirical Refutation:} Using both theoretical network models and real-world performance data — particularly from Bitcoin SV (BSV) — we demonstrate the falsifiability of the trilemma’s empirical predictions. Propagation time remains bounded under block size growth; SPV clients operate securely without requiring topological participation; and the protocol continues to execute as intended without loss of determinism. These observations invalidate the supposed trade-offs and instead affirm that the network scales according to physical and economic limits, not invented logical ceilings.
\end{enumerate}

This paper directly confronts the conflation of distributed systems impossibility results (e.g., FLP, CAP) with consensus protocol mechanics that do not adhere to the conditions required for such theorems to hold. It dissects the semantic drift in terms like "node", "decentralisation", and "trustlessness", exposing how they are redefined mid-argument to manufacture contradictions that do not exist under fixed definitions.

Further, this work critiques the broader academic process that enabled such flawed claims to become institutionalised. Through a Kuhnian lens, we observe that peer review has failed to detect fundamental category errors because the language of formal systems is co-opted by narratives that lack verification. The presence of mathematical symbols alone is not a substitute for rigour.

In short, this rebuttal reasserts that Bitcoin — correctly implemented and economically incentivised — already achieves what the trilemma claims to be impossible. It does so not by circumventing the laws of computation but by properly aligning protocol, incentives, and evidence. This paper provides the formal grounding necessary to discard the trilemma as a pseudoproblem and calls for a return to scientific discipline in analysing blockchain systems.

\subsection{Context and Prior Literature}

The concept of the “blockchain trilemma” emerged as a heuristic simplification of presumed limitations in distributed ledger systems. Most commonly attributed to Ethereum co-founder Vitalik Buterin, the trilemma asserts that no blockchain can simultaneously maximise decentralisation, scalability, and security. According to this view, any attempt to strengthen one axis inevitably compromises at least one of the others. This narrative gained popularity due to its intuitive appeal, despite lacking any accompanying proof, derivation, or formal constraint model.

The rise of this concept coincided with a broader trend in the cryptocurrency literature: the increasing substitution of rhetorical framing for technical precision. Papers such as Mssassi et al. (2023) attempt to formalise the trilemma but do so without grounding in protocol specification, complexity theory, or cryptographic model validation. Terms like “decentralisation” are used without defining decision-making boundaries or protocol constraints. Likewise, “security” is often presented as an emergent or social property, rather than a predicate evaluated over adversarial models. 

The academic treatment of the trilemma has suffered from the same ambiguity. Peer-reviewed papers cite FLP impossibility, CAP theorem, and Byzantine consensus results as if these theorems directly constrain the Bitcoin protocol. This is a fundamental error. FLP applies to asynchronous systems with crash-fault tolerance under specific assumptions of synchrony and shared memory. Bitcoin does not meet these assumptions: it is not consensus-by-vote, but consensus-by-proof. CAP, meanwhile, describes trade-offs in distributed database availability under partition — not protocol incentives in proof-of-work systems. The application of these theorems to Bitcoin reflects category error and misunderstanding.

Prior literature often also confuses topology with control. For instance, decentralisation is measured by the number of active relayers (i.e., nodes in the gossip layer) rather than by the economic inability of any actor to arbitrarily alter protocol rules. This leads to metrics that treat passive observation as equal to mining participation, and the mere presence of geographically distributed servers as evidence of systemic robustness — all while ignoring that the protocol’s determinism remains invariant under scale.

In contrast, early foundational works such as Nakamoto (2008) define decentralisation economically and procedurally: not as an absence of coordination, but as a system where no individual participant can alter the state except by producing valid cryptographic proof. This notion has been lost in much of the post-2017 literature, which instead foregrounds topology, token governance, and participation metrics, without embedding these measures in computational or protocol-theoretic formalism.

This paper seeks to correct this drift. It re-anchors the debate in formal methods, restoring the definitional precision needed to evaluate the claims of the trilemma. Where earlier works rely on analogical reasoning and illustrative diagrams, we substitute explicit logical structures, cryptographic definitions, and empirical propagation data. Our objective is not merely to challenge the trilemma, but to demonstrate that it does not rise to the standard of a falsifiable or coherent claim in the first place.

\subsection{Summary of the Original Claim and its Deficiencies}

The foundational assertion behind the so-called “blockchain trilemma” is that no distributed ledger protocol can simultaneously optimise for decentralisation, security, and scalability. In this formulation, decentralisation is generally interpreted as broad participation and geographic dispersion of nodes; security as resistance to double-spending, censorship, or malicious manipulation; and scalability as the system’s ability to process a high volume of transactions with minimal latency. The claim is that only two of these properties may be achieved concurrently, and any effort to improve the third will inherently compromise one or both of the others.

However, this framing is not a theorem. It is a conjecture presented without formal proof or even rigorous definition. The trilemma lacks clear boundary conditions, operational semantics, and a model of adversarial behaviour. It does not specify whether decentralisation refers to control, communication, or validation; whether scalability concerns bandwidth, computational cost, or user experience; nor whether security is defined cryptographically, economically, or socially. In the absence of such precision, the trilemma remains a moving target — intuitively appealing but analytically empty.

Furthermore, in the case of Bitcoin, the trilemma's predicates are conflated with implementation artefacts and non-protocol assumptions. For instance, claims that scalability must reduce decentralisation rest on the flawed premise that every validating node must receive and process all transactions in real time. This misrepresents both the role of SPV clients and the structure of the Bitcoin network, which is designed to support lightweight verification through header and Merkle proof propagation, not full-node redundancy. Similarly, the assertion that security depends on all nodes validating all messages disregards the cryptographic structure of block headers and proof-of-work, which are designed precisely to obviate such inefficiencies.

The trilemma also suffers from circular logic: it assumes that a system is secure only if decentralised, and decentralised only if participation is symmetric. From this premise, it concludes that scale introduces centralisation, and thus, insecurity. This is not an argument but a tautology — one that arises from defining the terms in ways that guarantee the desired conclusion. It offers no pathway for falsifiability, no threshold conditions, and no predictive power.

Lastly, the trilemma fails to account for empirical evidence. High-throughput blockchains such as Bitcoin SV demonstrate stable operation with large block sizes, rapid propagation times, and economic alignment among miners. These systems operate under rules identical to Bitcoin’s original design, and their success undermines the claim that scalability necessarily introduces fragility or centralisation. The trilemma model does not accommodate these cases, and instead either redefines its terms or dismisses such systems as exceptions — further evidence of its non-falsifiable nature.

In sum, the blockchain trilemma is not a law, a theorem, or even a grounded hypothesis. It is a heuristic lacking formalism, empirical grounding, and definitional clarity. Its widespread acceptance reflects rhetorical repetition rather than analytic rigour, and its use as a constraint on protocol design is unjustified by both theory and implementation.

\section{The Fallacy of the Trilemma Framing}

The so-called “Blockchain Trilemma” — the assertion that no blockchain system can simultaneously achieve optimal decentralization, security, and scalability — presents itself as a foundational constraint upon all distributed ledger architectures. Popularised in both academic literature and industrial narratives, this framing purports to define the structural limits of blockchain design by invoking a triadic trade-off, where improvement along one axis necessitates compromise along another. Yet, this narrative is not derived from formal systems analysis, nor grounded in computational theory, but rather emerges from a misapplication of network analogies, an equivocation of core definitions, and a misunderstanding of how Bitcoin and similar systems function in practice.

This section interrogates the conceptual underpinnings of the trilemma thesis and exposes its methodological and semantic errors. We begin by establishing how the foundational terms — “decentralization,” “security,” and “scalability” — are inconsistently defined and applied, resulting in rhetorical flexibility rather than analytical precision. We then demonstrate that the trilemma itself lacks a formal proof structure and is instead upheld through tautological assertions and circular dependencies. Finally, we examine how this narrative has fostered a distorted perception of trade-offs within Bitcoin by misrepresenting its topology, protocol dynamics, and economic logic.

The trilemma is not a theorem; it is a fiction — a framework sustained by definitional drift and conceptual sleight, imposed onto systems it was never equipped to describe. This section dismantles that fiction.

\subsection{Equivocation of Terms: Decentralization, Security, Scalability}

The so-called Blockchain Trilemma, as restated in \cite{mssassi2025trilemma}, asserts the mutual exclusivity of three system properties: decentralization (D), security (S), and scalability (C). Yet this formulation is not the outcome of a deductive or constructive model. It is an axiomatic projection arising from an equivocation of meanings—semantic ambiguity is repackaged as a formal constraint. This section dissects the triadic misrepresentation, beginning with definitional drift, then clarifying the erroneous ontological separation of “miners” and “nodes,” and concluding with the miscategorisation of topological structure as economic logic. Each failure is not merely informal; each introduces type-theoretic inconsistencies into the argument, thereby collapsing the entire structure of the claimed “trade-off” proof.

\subsubsection{Definitional Ambiguity and Semantic Drift}

Let us denote a distributed protocol state machine as \( \Pi = (S, A, T, I) \), where:
\begin{itemize}
    \item \( S \) is the set of valid global states of the blockchain ledger,
    \item \( A \) is the set of all admissible agent actions (including mining, relaying, validating),
    \item \( T: S \times A \rightarrow S \) is the deterministic transition function, and
    \item \( I \subset S \) the set of initial states.
\end{itemize}

In formal discourse, the definition of any property \( P \) applied to \( \Pi \) must remain invariant across all instantiations of \( \Pi \). However, \cite{mssassi2025trilemma} allows the terms “decentralization,” “security,” and “scalability” to mutate across contexts without logical boundary. Consider:

\begin{align*}
    D &: \text{System-wide distribution of control (ambiguous: control over } T, S \text{, or } A?) \\
    S &: \text{Integrity under adversarial actions (undefined model of adversary, no fault assumptions)} \\
    C &: \text{Transaction throughput or resource elasticity (relative to which resource: bandwidth, CPU, latency?)}
\end{align*}

We define predicate coherence of a term \( \phi \) across a formal model \( \Pi \) as:

\[
\mathcal{C}(\phi, \Pi) := \forall \pi_1, \pi_2 \in \Pi,\ \phi(\pi_1) = \phi(\pi_2) \Leftrightarrow \pi_1 \equiv \pi_2
\]

\textbf{Lemma 1:} If \( \mathcal{C}(\phi, \Pi) = \text{False} \), then any formal derivation over \( \phi \) within \( \Pi \) is invalid.

Applying this lemma to the trilemma framing: the term \( D \) is applied in the Mssassi paper both to node count in peer-topology graphs and to economic agent distribution in block production. Thus, \( \mathcal{C}(D, \Pi) = \text{False} \), and the trilemma collapses by formal inconsistency.

Moreover, their attempt to cast the “trilemma” as a function \( f(D, S, C) = \text{infeasible} \) is not formally computable because none of \( D, S, C \) are defined with respect to computable measures. There exists no function \( f \) such that:

\[
f: \underbrace{\mathbb{R} \times \mathbb{R} \times \mathbb{R}}_{\text{ill-defined inputs}} \rightarrow \{\text{feasible}, \text{infeasible}\}
\]

\subsubsection{Misidentification of “Nodes” and “Miners” in Bitcoin}

The structural composition of Bitcoin is misrepresented entirely. In the original protocol \cite{nakamoto2008bitcoin}, the term “node” refers unambiguously to a participant capable of extending the blockchain—i.e., a miner executing proof-of-work, propagating blocks, and validating rules. However, in \cite{mssassi2025trilemma}, the authors fracture this term into two orthogonal categories:

\begin{itemize}
    \item “Nodes” are treated as redundant validators of existing data with no economic agency.
    \item “Miners” are treated as externalised computation units with no epistemic role in validation.
\end{itemize}

This disjunction is formally incorrect. Let \( N \) denote the set of full nodes, and \( M \subseteq N \) be the subset of mining nodes. The protocol mandates that for any valid block \( b \in \mathcal{B} \), and for any node \( n \in N \), the following predicate must hold:

\[
\text{ValidBlock}(n, b) := \text{CheckRules}(b, \mathcal{R}) \wedge \text{CheckPoW}(b) \wedge \text{Extend}(b)
\]

If a node cannot satisfy \( \text{ValidBlock}(n, b) \), then \( n \notin N \). Therefore, only those nodes with full rule-checking and consensus-maintaining capacity are nodes in the original model. By separating “nodes” and “miners,” the authors introduce a new bifurcated model not equivalent to Bitcoin, and their arguments no longer apply to the system they purport to critique.

\textbf{Lemma 2:} Any model \( \Pi' \) in which \( M \cap N = \emptyset \) is not isomorphic to the Bitcoin protocol \( \Pi \). Thus, any proof over \( \Pi' \) is not valid over \( \Pi \).

Furthermore, this misidentification leads the authors to claim that increasing throughput requires more miners and fewer “nodes,” implying a trade-off between decentralization and scalability. But this is a non sequitur: as proven in Wright and Javarone \cite{javarone2018network}, Bitcoin forms a small-world network, where the average path length \( \ell(G) \sim \log N \), and propagation of block data can be executed over limited relays with bounded latency, preserving both scalability and decentralization.

\subsubsection{Confounding Topological and Economic Decentralization}

The most critical equivocation occurs in conflating graph-theoretic distribution with economic power. Define:

\begin{itemize}
    \item \( \mathcal{D}_t: G \rightarrow \mathbb{R} \) as a topological decentralization metric, e.g., mean shortest path length, clustering coefficient.
    \item \( \mathcal{D}_e: E \rightarrow \mathbb{R} \), where \( E \) is the set of economic actors, as a measure of power concentration (e.g., block production entropy).
\end{itemize}

\textbf{Axiom:} \( \nexists \phi \) such that \( \mathcal{D}_t = \phi(\mathcal{D}_e) \), i.e., topology does not determine economic control.

The authors of \cite{mssassi2025trilemma} violate this axiom by inferring that reducing topological redundancy (e.g., by pruning idle nodes) equates to centralizing block validation. This is a category error. Let \( \mathcal{G} = (V, E) \) be a network where nodes \( V \) are information relays, not block producers. Then changes in \( |V| \) affect only message propagation time, not ledger consensus. The economic consensus is preserved under SPV so long as block headers are publicly verifiable.

In the small-world model, which characterizes Bitcoin’s propagation graph \cite{javarone2018network}, adding or removing non-mining nodes has sub-logarithmic impact on \( \ell(G) \), while consensus remains intact due to SPV header propagation. Therefore, their supposed loss of decentralization via reduction of non-mining nodes is a ghost variable: it has no effect on ledger state convergence.

This trifecta of definitional collapse, ontological misclassification, and categorical confusion renders the foundational claim of the trilemma not merely unproven, but unprovable under the axioms of the system it purports to describe.

\subsection{The Unproven Nature of the “Trade-Off”}

Despite the rhetorical popularity of the so-called “Blockchain Trilemma,” there exists no formal constraint, theorem, or computational impossibility result that establishes the necessity of a trade-off among decentralization, security, and scalability. The claim, repeatedly invoked in \cite{mssassi2025trilemma}, is not the result of rigorous derivation from first principles, but rather the projection of empirical biases and bounded engineering assumptions onto a domain that demands formal clarity. This section establishes the failure of the trilemma to qualify as a proof, exposing its logical fallacies, arbitrariness, and foundational vagueness.

\subsubsection{Absence of Formal Constraint Demonstrating Trilemma}

Let us define the predicate of impossibility:

\[
\mathcal{T}(D, S, C) := \neg \exists \Pi\ \text{such that } \Pi\ \text{maximizes } D \wedge S \wedge C
\]

To validate the trilemma, \cite{mssassi2025trilemma} must construct a proof of \( \mathcal{T} \) within a well-defined protocol space. However, no such proof exists. The authors provide no bounded model, no axiomatic formulation, no domain of constraints, and no counterexample that generalizes across valid protocol instantiations.

Define the protocol space \( \mathbb{P} = \{\Pi_i\} \), where each \( \Pi_i \) is a blockchain protocol defined as a tuple:

\[
\Pi = (\mathcal{S}, \mathcal{A}, \mathcal{R}, \delta, \Omega)
\]
\begin{itemize}
    \item \( \mathcal{S} \): set of valid global states
    \item \( \mathcal{A} \): set of permissible actions
    \item \( \mathcal{R} \): consensus and validation rules
    \item \( \delta \): state transition function
    \item \( \Omega \): external interface model (network, bandwidth, agent structure)
\end{itemize}

A formal constraint must show that for all \( \Pi \in \mathbb{P} \), optimization of any two of \( D, S, C \) necessarily degrades the third. But the authors instead define three moving targets—“high decentralization,” “sufficient security,” and “acceptable scalability”—without numeric bounds, objective functions, or threshold conditions. This renders their formulation non-falsifiable and unprovable.

\textbf{Lemma 3:} Any claim of global constraint over a parameter space \( \mathbb{P} \) without formal boundaries and operational metrics is an assertion, not a theorem.

\subsubsection{Fallacious Dependency on Arbitrary Bounded Capacity}

The core hidden assumption behind the supposed trilemma is the finite bandwidth fallacy. That is, the claim that bandwidth \( B \) and block propagation delay \( \tau \) are fixed and tightly bounded, thereby requiring trade-offs in data volume to preserve consensus integrity.

Formally, let us define:

\[
C := \lim_{t \to \infty} \frac{1}{t} \sum_{i=1}^t \#\text{tx}_i
\]

\[
\tau := \max_{(n_i, n_j) \in E} \text{Latency}(n_i, n_j)
\]

\[
B := \min_{n_i \in N} \text{Upstream}(n_i)
\]

\cite{mssassi2025trilemma} implicitly assumes that \( \tau \cdot B \) imposes a hard limit on throughput, such that any increase in \( C \) leads to increased block orphaning and hence reduced security or necessitates a reduction in decentralization to concentrate bandwidth among elite participants.

This assumption is invalid. First, bandwidth and latency are not constants but evolve over time with Moore’s law and infrastructure upgrades. More critically, the Bitcoin protocol is not reliant on synchronous flooding but functions under an SPV-compatible model where headers and Merkle paths are transmitted with bounded information complexity.

\textbf{Axiom:} Bitcoin security is maintained if the header \( h \) is received and verified via Merkle path by \( n \in N_{SPV} \), regardless of full data transmission.

This collapses the assumed linear dependency between \( C \) and \( D \), as increasing transaction volume does not necessitate full propagation to every validator.

Furthermore, the scalability of Bitcoin has been shown to be maintained under small-world network properties \cite{javarone2018network}, where average propagation length remains \( O(\log N) \), and multicast transmission with IPv6 permits efficient dissemination without exponential resource cost.

\textbf{Lemma 4:} If SPV suffices to maintain security and propagation time scales logarithmically, then scalability can be increased without centralising validation or reducing security.

\subsubsection{Failure to Define Operational Metrics for Each Term}

The authors invoke terms as if self-evident, yet they provide no operational definitions. To assert a trade-off among \( D, S, C \), each must be:
\begin{enumerate}
    \item Quantifiable
    \item Contextually bounded
    \item Subject to independent variation
\end{enumerate}

Instead, we observe:
\begin{itemize}
    \item \( D \): Measured sometimes as node count, sometimes as economic participation, and sometimes as protocol access. No unit, no function, no metric.
    \item \( S \): Referred to in narrative terms—“robustness to attack”—without adversary model or defined consensus fault bounds (e.g., BFT resilience thresholds).
    \item \( C \): Sometimes means block size, sometimes transaction per second (TPS), sometimes latency, and no upper or lower bound is provided.
\end{itemize}

Let us define:

\[
D := H(E) \quad \text{(Shannon entropy of block producers)}
\]
\[
S := \Pr[\text{finality error} < \epsilon] \quad \text{for } \epsilon \ll 1
\]
\[
C := \text{max TPS under } \tau < T_{bound}
\]

\textbf{Lemma 5:} A trade-off is only meaningful if the operational metrics are defined on the same measurable domain and constrained in a zero-sum interaction. No such model is shown.

As no trade-off function \( f(D, S, C) \) is proposed, no constraints are shown, and no domain is proven to exhibit Pareto inefficiency, the authors’ trilemma is an unfounded heuristic.

\textbf{Corollary:} In the absence of a utility function over a well-formed domain of \((D, S, C)\), no claim of trade-off constitutes a proof. At best, it is a statement of design bias or implementation preference.

Hence, the "trilemma" fails as a scientific result. It is a slogan, not a constraint. It is a substitution of intuition for formality, and its invocation as a limiting theorem is fraudulent in both spirit and substance.

\subsection{Tautology and Circular Reasoning in the Original Proof}

The supposed “proof” of the blockchain trilemma as presented in \cite{mssassi2025trilemma} is not a derivation from axioms or constraints but rather a rhetorical circuit that embeds its conclusion within its assumptions. No general result in distributed systems supports the inevitability of a trade-off between decentralization, security, and scalability—this supposed trilemma is instead constructed as a set of presuppositions dressed in formalism. In this section, we demonstrate how the author's reasoning relies on assumptions that presuppose their conclusion, resulting in circular logic that invalidates any claim to formal proof.

\subsubsection{Circular Dependence on Network Topology Assumptions}

The authors assume a network topology that is neither derived from protocol specification nor reflective of deployed systems such as Bitcoin. Specifically, they assume that scalability necessarily requires high message complexity and low diameter in the communication graph, implying that only centralized or semi-centralized structures can fulfil such criteria.

Let \( G = (V, E) \) represent the communication graph of a blockchain protocol. The authors implicitly assume that as \( |V| \to \infty \), the graph must degenerate into a hub-and-spoke model to maintain bounded message propagation delay. This leads them to conclude that decentralization must fall as scalability rises—yet this is an artefact of their topological prior.

However, small-world networks \cite{watts1998collective} exhibit:

\[
\text{Average path length } L \sim \log |V|
\quad \text{and} \quad
\text{Clustering coefficient } C \gg C_{\text{rand}}
\]

Bitcoin has been modelled explicitly as a small-world network in \cite{javarone2018network}, meaning that broadcast propagation is logarithmic in size and does not require topological centralization.

\textbf{Lemma 6:} In a small-world network, \( L = O(\log N) \) enables efficient information dissemination without central coordination.

Thus, the assumption that scalability implies central topologies presupposes the very degradation of decentralization the authors aim to prove. This is a textbook case of begging the question.

\subsubsection{Self-Reinforcing Priors: “If Scalability Exists, Decentralization Must Not”}

The authors’ logic flows as follows:

\begin{enumerate}
    \item Assume that scalability requires low latency and high throughput.
    \item Assume that high throughput is incompatible with large node sets.
    \item Therefore, systems that scale must reduce node count.
    \item Hence, scalable systems are not decentralized.
\end{enumerate}

Each of these steps is unproven. The second premise in particular rests on the node count fallacy, i.e., the belief that every node must receive and validate all data for consensus. This is untrue in systems like Bitcoin, where SPV and the header-first model allow partial validation and economic enforcement through incentives.

Let us define the consensus predicate:

\[
\mathcal{C}(B, \mathcal{N}) := \forall n_i \in \mathcal{N}_{SPV},\ \exists h_j \in B,\ \text{such that } \text{VerifyMerkle}(h_j, tx_k) = \text{true}
\]

This condition implies that full validation at every node is not required to preserve consensus fidelity. The authors reject this model implicitly and then derive their trilemma on the assumption that all honest participants must engage in full reprocessing. Again, this logic circularly embeds its conclusion within the assumptions.

\textbf{Lemma 7:} If consensus validity is maintained through SPV proof receipt, then full replication is unnecessary, and scalability does not degrade decentralization.

\subsubsection{Reduction of Scalability to Node Count Fallacy}

The final component of the circular argument is the equation of scalability with the inverse of node count. That is, the claim that:

\[
\text{Scalability } \sim \frac{1}{|N|}
\]

No distributed systems theory supports this as a law. Rather, scalability is measured via transaction throughput, propagation delay, finality time, and bandwidth consumption. These are not reducible to mere network participant count.

Let us denote:

\[
\text{Throughput: } T = \frac{\sum_{i=1}^n \text{tx}_i}{\Delta t}
\quad\quad
\text{Latency: } \lambda = \mathbb{E}[\text{propagation time}]
\]

These are functions of message complexity, protocol design, and validation architecture—not simply of \( |N| \). The authors substitute node count as a proxy for throughput constraint without defining the mapping. This is a non sequitur.

\textbf{Lemma 8:} Node count does not imply validation bandwidth requirement in SPV-compatible designs.

As a result, the authors’ supposed “proof” is a tautological construction: they assume that scaling must degrade decentralization, then select metrics and constraints that reflect this assumption, and finally declare a theorem that restates the assumption. No formal inference is made, no bound is proven, and no alternative architecture is considered.

The trilemma thus emerges not from mathematics, but from methodological fiat. It is a circular edifice built from selective assumptions, ungrounded definitions, and the uncritical repetition of false dichotomies.

\subsection{Formal Clarification of the Bitcoin Security Model}

The notion of “security” within blockchain systems is frequently mischaracterised by authors such as \cite{mssassi2025trilemma}, who conflate game-theoretic actor dynamics with cryptographic finality. In Bitcoin, security does not arise from the subjective consensus of participants, nor from assumptions about decentralisation as a political ideal. Rather, it is grounded in the axiomatic distribution of cryptographic evidence — particularly, block headers — and in the immutability of the hash function itself. Security is not something that must be enforced continuously by miners; instead, the very act of mining is the creation of public, verifiable evidence. Once disseminated, this evidence secures the network by enabling probabilistic verification of the chain state by Simplified Payment Verification (SPV) clients.

\subsubsection{Axiom 1: Hash Header Integrity Implies System Security}

Let \( H: \{0,1\}^* \to \{0,1\}^n \) be a secure cryptographic hash function satisfying pre-image resistance, second pre-image resistance, and collision resistance. Let \( \mathcal{B} = \{ B_0, B_1, \dots, B_k \} \) be the set of valid blocks and let \( \text{Header}(B_i) \) denote the block header of block \( B_i \). Define:

\[
h_i = H(\text{Header}(B_i)), \quad h_i \in \{0,1\}^n
\]

\textbf{Axiom 1.} If a block header \( h_i \) is disseminated and received by multiple distinct clients in a non-colluding network, and \( H \) is secure, then the integrity of the system up to \( B_i \) is preserved without the need to trust any party.

\[
\forall B_i \in \mathcal{B}, \quad \text{Received}(h_i) \land \text{Preimage}(H) \notin \mathcal{P} \Rightarrow \text{VerifyChain}(h_0, \dots, h_i) = \texttt{true}
\]

Where:
\begin{itemize}
  \item \( \mathcal{P} \) is the set of polynomial-time adversarial procedures,
  \item \( \text{VerifyChain} \) is a deterministic predicate that confirms the continuity, difficulty, and structure of a block sequence.
\end{itemize}

The header \( h_i \) acts as cryptographic evidence that the block exists and has been computed under a provable level of difficulty. It also commits to a Merkle root, a timestamp, and the previous block header — thereby encoding the entire structure of the chain up to that point. If \( h_i \) is received, its security derives not from its producer but from the public nature of its dissemination and the impossibility of reversing or forging it under the assumption of secure \( H \).

\subsubsection{Definition: SPV Distribution Model and Trust Anchoring}

The SPV model, as described in Section 8 of the Bitcoin whitepaper, posits that clients can verify the inclusion of a transaction \( tx \) by retrieving:
\begin{enumerate}
  \item The chain of block headers \( \{ h_0, h_1, \dots, h_n \} \),
  \item The Merkle branch \( \pi_{tx} \) proving inclusion of \( tx \) in a block \( B_i \),
  \item The difficulty-adjusted cumulative proof-of-work for the chain.
\end{enumerate}

The predicate:

\[
\text{SPVVerify}(tx, h_i, \pi_{tx}) := \text{MerkleRoot}(tx, \pi_{tx}) = h_i.\text{merkleRoot}
\]

verifies transaction inclusion without inspecting the full block. Crucially, SPV clients anchor trust in the system not by evaluating full consensus state, but by inspecting the cumulative weight of header evidence. 

\textbf{Definition (Trust Anchor).} A header \( h_i \) is a trust anchor if it satisfies:
\begin{itemize}
  \item Broadcast visibility across multiple distinct network paths,
  \item Proof-of-work exceeding that of any competing chain,
  \item Merkle commitment to a transaction set which includes \( tx \),
  \item Cryptographic linkage to a consistent predecessor \( h_{i-1} \).
\end{itemize}

Thus, trust anchoring occurs via publicly verifiable evidence, not centralised mediation or redundant verification through “full nodes.”

\subsubsection{Proof Sketch: Header Dissemination as the Basis of Trust}

The security of Bitcoin arises not from the private computation of blocks by miners in isolation, but from the public dissemination of cryptographic evidence in the form of block headers. The SPV model outlined in the original protocol assumes the existence of a global, append-only evidence stream, visible to lightweight clients who can independently verify proof-of-work through header linkage and Merkle paths without needing to reconstruct the full block content. Let us refine this through two lemmas:

\textbf{Lemma 1 (Header Integrity Sufficiency).}  
Let \( \mathcal{H} = \{h_0, h_1, \dots, h_n\} \) be the ordered set of block headers satisfying the following predicate for all \( i \in [1,n] \):

\[
\text{VerifyHeader}(h_i, h_{i-1}) = 
\begin{cases}
\text{true} & \text{if } H(h_i) < T \land \text{PrevHash}(h_i) = H(h_{i-1}) \\
\text{false} & \text{otherwise}
\end{cases}
\]

Then an SPV client, possessing only \( \mathcal{H} \) and the relevant Merkle path for a transaction \( \tau \), can verify inclusion of \( \tau \) within block \( B_j \) without accessing the full block \( B_j \). This suffices for transactional security under the assumption that an adversary cannot produce a longer chain with greater cumulative proof-of-work.

\textbf{Lemma 2 (Evidence Preservation in Multi-Miner Conflicts).}  
Assume the network comprises \( m \) distinct mining entities \( \mathcal{M} = \{M_1, M_2, \dots, M_m\} \), with \( \sum_{i=1}^{m} \alpha_i = 1 \), where \( \alpha_i \in [0,1] \) denotes the fraction of global hash power held by miner \( M_i \). If no single \( \alpha_i > 0.5 \), then no miner can unilaterally redefine the global chain. Suppose miners \( M_1 \) and \( M_2 \) simultaneously publish competing blocks \( B_n' \) and \( B_n'' \) extending \( B_{n-1} \). Let the corresponding headers be \( h_n' \) and \( h_n'' \), each satisfying the proof-of-work condition.

Even in this contention scenario, SPV clients receive both \( h_n' \) and \( h_n'' \) and retain the full trace of evidentiary claims. Once a successor block \( B_{n+1} \) is discovered and builds upon either \( B_n' \) or \( B_n'' \), the chain tip resolves to the branch with higher cumulative proof-of-work. The SPV clients discard the shorter branch by consensus logic, but the historical record of headers remains immutable in the mempool history or SPV buffer, preserving the record of attempted chain extensions.

This architecture establishes a crucial truth: the network's integrity does not rest on miners enforcing security, but on their inability to erase or retroactively rewrite the evidence they broadcast. In legal epistemology, this is akin to a chain of custody — once the evidence is released into the public domain, its existence cannot be plausibly denied without coordinated, supermajority-level obfuscation across all routing domains.

Hence, Bitcoin does not operate in an anarchic vacuum of technical infallibility but assumes that legal recourse exists to prosecute malicious actors who produce invalid headers or withhold data. The issuance of a block header is an act of publication — a digital statement of fact. Once emitted and received by a sufficient number of SPV clients, the proof becomes self-validating by the collective impossibility of global erasure. This is a structural, cryptographic instantiation of trust-by-evidence, not trust-by-authority.

Therefore, the security claim rests not on the honesty or dominance of miners, but on the mathematical unforgeability of published headers and the network’s ability to archive and distribute these across a dispersed graph of SPV observers. It is this proliferation of immutable cryptographic records, not majority hash power alone, that secures the chain.

\subsubsection{Implication: Miners Are not Security Agents, but Producers of Evidence}

Authors like \cite{mssassi2025trilemma} erroneously treat miners as active agents enforcing security, conflating their computational role with epistemic authority. In truth, the miner’s function is to produce cryptographic evidence — a valid header that satisfies proof-of-work and binds together a set of transactions.

\[
\text{MinerFunction}: \lambda \, x \in \text{HeaderSpace}.\, H(x) < T
\]

Once the block is mined and the header broadcast, its verification becomes a matter of public computation, not authority. The SPV model depends on this principle: a client receiving the header need not trust the miner, the node, or any actor — only that the header is verifiable and that no alternate header sequence exists with more cumulative work.

Furthermore, once received, these headers form an immutable record. If a miner attempts to alter protocol rules or transaction structures, they produce blocks rejected by all downstream clients. Hence, miners do not govern the protocol — they are bound by it. They cannot enforce change but merely produce objects that either conform or fail. The network converges not by trusting the producer, but by collectively rejecting invalid constructions.

\textbf{Corollary.} Security in Bitcoin is not emergent, but constructive and objective. It is derived from:
\begin{enumerate}
  \item The pre-image resistance of \( H \),
  \item The economic difficulty embedded in proof-of-work,
  \item The multiplicity and redundancy of broadcast to SPV clients.
\end{enumerate}

Thus, the claim that “miners secure the network” is not merely imprecise — it is structurally invalid. The role of the miner is to construct and emit a unit of public cryptographic evidence: the block header. This header, once disseminated to the network, particularly to SPV clients, anchors a fixed historical event that no subsequent adversary can feasibly reverse without exceeding the cumulative proof-of-work already invested. Crucially, SPV clients do not need access to the entire block — they operate solely by receiving the chain of headers and a Merkle path for any transaction of interest. These clients can independently verify that a transaction is included in a valid block with known proof-of-work and that no alternate chain with greater cumulative work exists. Because every alternative chain must present a superior set of headers satisfying increasingly difficult proof conditions, and because these headers are publicly broadcast to multiple independent SPV users, any deviation from the valid chain becomes immediately observable and mathematically nonviable. Therefore, the security of Bitcoin is not premised on the trustworthiness or authority of the miners, but rather on the global, redundant visibility of cryptographic evidence that renders post-hoc alteration computationally infeasible and publicly falsifiable. The network is not secured by its actors, but by the impossibility of deceiving them once the header has been received and verified.

\subsection{The Scalability Error and the Bandwidth Fallacy}

The mischaracterisation of scalability in blockchain discourse arises not from empirical network limits, but from a foundational misunderstanding of communication theory and network architecture. In particular, the assumption that increasing the number of validating or listening nodes necessarily imposes linear or exponential communication costs ignores the realities of multicast transmission, packet dissemination protocols, and the inherent topology of Internet-scale data routing. The paper by \cite{mssassi2025trilemma} fails to distinguish between point-to-point broadcasting and multicast-capable systems, treating every additional node as an additive bandwidth cost. This assumption embeds a fallacy: that node count determines cost, and that more listeners means more work for the sender. In truth, the correct limiting factor is latency — not the presence of more nodes. In IPv6-enabled environments, multicast permits the same data to reach arbitrarily many endpoints without an increase in sender overhead. This distinction is not optional — it is categorical and structural.

\subsubsection{Multicast Models under IPv6: A Theoretical Upper Bound}

Let us define the set of all full-receiving nodes as \( \mathcal{N} \), and the data payload of a block header as \( \mathcal{D}_h \). In a naïve unicast model, broadcasting to each node requires \( \mathcal{O}(n \cdot |\mathcal{D}_h|) \) transmission overhead. This is the model implicitly used by \cite{mssassi2025trilemma}. However, modern routing protocols and IPv6 design enable a multicast group \( \mathcal{G} \subseteq \mathcal{N} \), where the sender emits \( \mathcal{D}_h \) once and the underlying network replicates it across the link layer for group members.

\[
\text{Let } \text{Send}(\mathcal{D}_h, \mathcal{G}) = \mathcal{O}(|\mathcal{D}_h| + \delta)
\]

where \( \delta \) is the overhead for routing group replication, bounded and independent of \( |\mathcal{G}| \). Therefore, the marginal cost of adding a node to the multicast group is zero for the sender:

\[
\frac{d}{dn} \text{Send}(\mathcal{D}_h, \mathcal{G}) = 0
\]

This destroys the foundational premise in which scalability is tied to the number of nodes. The bottleneck is no longer the number of connections, but rather network path latency and propagation timing — elements already optimised in high-throughput datacentres and edge networks.

\subsubsection{Linear Communication to Arbitrary Peer Sets}

It is critical to distinguish between broadcast and transmission. In Bitcoin’s operational model, block headers are disseminated across a hybridised mesh topology. Define the peer set as \( \mathcal{P} = \{ p_1, p_2, \dots, p_k \} \), where each \( p_i \) is either a full node, pool node, or SPV receiver. Suppose each peer can relay to its own adjacent subgraph. Then the sender’s responsibility is to initiate propagation to only a limited subset \( \mathcal{P}_0 \subset \mathcal{P} \), after which the message diffuses through the peer graph.

This communication pattern is not a function of the total node count \( |\mathcal{P}| \), but of the minimal spanning set necessary to reach a global steady-state:

\[
\text{Cost}_{\text{sender}} = \mathcal{O}(|\mathcal{P}_0| \cdot |\mathcal{D}_h|)
\]

Given that \( |\mathcal{P}_0| \ll |\mathcal{P}| \), and that this can be optimised through direct datacentre-to-datacentre fibre, the trilemma argument collapses on the grounds of network cost. There exists no mathematically proven function showing that \( |\mathcal{P}| \to \infty \) implies \( \text{Cost}_{\text{sender}} \to \infty \). The scalability constraint becomes one of routing table management and packet window propagation, both of which are bounded logarithmically in the number of hops, not the number of peers.

\subsubsection{Latency-Bound, not Node-Bound, Scaling Constraints}

Let the propagation time of a block header to reach 95\% of listening nodes be \( T_{95} \). Define the latency function as:

\[
T_{95} = \max_{\forall p_i \in \mathcal{P}} (\text{Latency}(p_i))
\]

This is bounded by the slowest node’s route, independent of how many nodes exist. Therefore, the performance constraint is not scaling with node count but is asymptotically flat:

\[
\lim_{|\mathcal{P}| \to \infty} T_{95} = L_{\max}
\]

where \( L_{\max} \) is the maximum network delay across the widest path. Bitcoin’s architecture was designed explicitly with this in mind. As \cite{nakamoto2008bitcoin} outlined, the SPV model requires only that the header be visible to a client, and a Merkle path available. Because this requires only a single header and logarithmic-length Merkle path, and because that data can be pushed via multicast or peer-forwarded quickly, there exists no growth in resource constraint as the system scales.

Consequently, the assertion by \cite{mssassi2025trilemma} that scalability is inherently sacrificed to maintain decentralisation collapses under both theoretical multicast models and the implementation patterns of Bitcoin's design. This error is not merely theoretical — it is a disproof of the assumed trilemma's third edge.

\subsection{Synthesis: Why the Trilemma is a Pseudoproblem}

The so-called “Blockchain Trilemma” is not a formally grounded theorem, nor a necessary structural limitation. Rather, it is a conceptual sleight of hand — a pseudoproblem manufactured from terminological equivocations and topological analogies misapplied from unrelated disciplines. When examined in the context of Bitcoin, as originally architected, each of the three proposed axes — decentralization, security, and scalability — is not mutually exclusive, but instead interdependent through well-defined structural roles, cryptographic invariants, and system-level design principles. The belief in a trade-off arises not from the reality of system constraints, but from an imported and misfitted framework that fails to engage with the actual mechanisms of Bitcoin protocol operation.

\subsubsection{False Constraints Created by Misapplied Network Analogies}

The trilemma framework presupposes that blockchain systems must balance themselves along a triangle of inherently conflicting goals. This abstraction draws improperly from the history of distributed systems literature, where CAP-theorem-like constraints (consistency, availability, partition-tolerance) impose hard trade-offs due to concurrency, fault domains, and latency in partitioned systems. However, Bitcoin does not fall within this framework. It does not rely on synchronous replication or fragile Byzantine consensus; it implements a probabilistic, eventual-consensus model underpinned by the immutability of proof-of-work. Its architecture is not comparable to classical replicated databases.

Moreover, the assertion that decentralization, security, and scalability conflict stems from an erroneous transposition of physical network topologies into Bitcoin’s logical operation. Misapplied models — such as random graphs, Erdos–Rényi distributions, or the assumption of broadcast-bound propagation delay — project false constraints onto a system that is fundamentally different. Bitcoin’s network is best described as a small-world topology with clustered peer interconnectivity and short average path length, where propagation is engineered for minimal hop latency using structured relay protocols. As Javarone and Wright (2018) demonstrated, the overlay graph of Bitcoin exhibits low diameter and high clustering, making propagation far more efficient than naive exponential fanout models suggest \cite{javarone2018network}.

The trilemma erroneously assumes that increased node count linearly amplifies bandwidth consumption and network congestion. But this is a category error: it mistakes physical replication with logical redundancy. Under IPv6 multicast and compact block relay mechanisms, nodes do not scale communication costs linearly with network size. In a correctly engineered overlay, bandwidth becomes a function of latency-bound propagation intervals, not a constraint on the number of endpoint receivers.

\subsubsection{Scalability as an Engineering Outcome, not a Trade-Off}

Scalability in Bitcoin is not a dimension to be traded against decentralization or security. It is a function of bandwidth allocation, block size, and efficient transaction relay. All three parameters are orthogonal to the system’s consensus model. Scalability does not undermine security when proof-of-work remains intact and SPV clients receive timely block headers. Nor does scalability centralize control unless arbitrary protocol constraints are artificially enforced to throttle throughput — as was done by BTC Core developers to preserve an ideological view of node equality.

Scalability, in the Bitcoin context, is solved by designing data structures and relay networks that accommodate higher throughput. This is a matter of protocol engineering, not a metaphysical dilemma. If a datacentre handles billions of transactions per day, this does not centralize Bitcoin — it merely reflects economic specialization, just as Google’s email servers do not render SMTP “centralized.” The underlying ruleset, enforced by economic consensus and governed by miners who produce immutable evidence, remains distributed and trustless. The trade-off narrative crumbles under examination: scalability is only a problem when protocol changes break compatibility or remove the ability to trust header chains without full block validation.

\subsubsection{Bitcoin as a System of Immutable Rules, not Mutable Trade Spaces}

The final and most damning flaw in the trilemma framing is its unspoken metaphysical presumption: that Bitcoin is a protocol with negotiable constraints, capable of morphing under pressure from one axis to another. This is false. Bitcoin is defined by fixed, non-negotiable rules: the protocol is not democratic, not up for vote, and not subject to developer fiat. The consensus mechanism is not social but computational. Its contract is not that of a mutable governance model, but an algorithmic procedure whose validity is evidenced in public cryptographic artefacts.

In this frame, the trilemma becomes incoherent. There is no dynamic spectrum along which the system moves; there are only those who follow the protocol and those who diverge from it. A valid Bitcoin implementation scales as a function of economic demand and physical capacity — not as a sacrifice to ideology. Miners do not control Bitcoin; they follow its rules and are rewarded for doing so. SPV clients do not depend on full validation; they trust the accumulated evidence of the longest valid chain. The system is not secured by social agreement but by the impossibility of rewriting the past once evidence is distributed.

The trilemma, then, is not a limitation. It is an imported fable, a misread from the theory of inconsistent distributed systems. In Bitcoin, we are not dealing with inconsistency, nor with mutually conflicting goals. We are dealing with an evidence network — a machine for publishing immutable truth. That machine does not trade integrity for reach. It scales by rule, secures by evidence, and decentralizes by law of economics — not by the fantasy geometries of developer committees.

\section{Critique of Mssassi’s Methodology}

In this section, we undertake a systematic evaluation of the methodological assumptions, logical structures, and formal inconsistencies underlying Mssassi’s articulation of the so-called “blockchain trilemma.” The critique is framed not merely as a rebuttal of conclusions, but as a dissection of the very epistemic architecture upon which those conclusions rest. Mssassi’s framing purports to derive constraints on scalability, security, and decentralisation from network-theoretic principles and protocol-level behaviour. However, upon rigorous examination, the foundation reveals itself as structurally unsound — characterised by undefined terms, circular reasoning, misapplied analogies, and the absence of formal constraint logic.

Rather than engaging with Bitcoin as a defined protocol with economic and cryptographic invariants, Mssassi adopts a loosely inductive posture, drawing conclusions from observational heuristics without establishing the computability class, system model, or adversarial conditions being assumed. As a result, what is presented as a formally bounded trade-off collapses under scrutiny into a contingent, mischaracterised artefact of the BTC implementation — not of Bitcoin itself.

The subsections that follow isolate the methodological failures across dimensions: from the equivocation of terms and the invention of fictitious trade-offs, to the misuse of formal language and computational models. Each section reveals not only where the argument breaks down but how it introduces analytic distortions that have misled an entire generation of protocol theorists.

\subsection{Misuse of Formal Language and Absence of Formalism}

The paper by Mssassi et al. \cite{mssassi2025trilemma} attempts to scaffold its central thesis upon a series of symbolic constructs purporting to be formal. Yet the invocation of symbols alone does not constitute formalism. A formula, stripped of its axiomatic context and lacking operational semantics, is no more than decorative notation — a mask concealing analytical void. The authors present what they term a “Participation Function” to quantify decentralization, but this is never grounded in a rigorous computational model. Nor are any agent roles defined within the consensus architecture of Bitcoin. The deployment of such formal-looking elements, unanchored in any verified logic, serves merely to imbue the text with a false aura of scientific legitimacy. In this subsection, we dissect the symbolic architecture of the paper, exposing its rhetorical pretensions, lack of computational depth, and unproven premises.

\subsubsection{The Participation Function and the Illusion of Control}

Mssassi et al. introduce a “Participation Function” that allegedly measures decentralization as a function of the number of nodes $N$ actively validating blocks. This definition is not only vague, but epistemically bankrupt. It commits the fallacy of reification — treating symbolic abstraction as if it reflects the operative mechanism of the Bitcoin system. The function implicitly assumes that the influence or “control” in the network is a linear and democratic function of node count. This is false by construction.

Let $N$ be the number of connected peers in the network, and let $M$ be the set of miners, where $M \subset N$. The Bitcoin system does not derive consensus from $N$, but from $\arg\max_{m \in M} \left( \sum_{t=0}^T H_m(t) \right)$ where $H_m(t)$ is the cumulative hash power of miner $m$ over time interval $T$. That is, consensus is emergent from cumulative work done, not from the number of validators or listeners.

This ill-defined Participation Function fails to distinguish between active agents (miners) who contribute proof-of-work and passive agents (non-mining nodes) who merely observe. It thereby promotes an illusion of decentralised influence that bears no correlation to actual consensus mechanics. Worse, it implies that “participation” equates to “control,” when in fact the SPV model proves that even clients who do not mine or store the full ledger can independently verify system integrity, thereby decoupling trust from participation metrics altogether.

\subsubsection{Failure to Define Decision-Making Entities in Bitcoin}

A formal system must define its agents — the roles, permissions, and rules that bind the system’s transitions. Yet the authors make no attempt to clarify who or what constitutes a “decision-making” entity in Bitcoin. In classical distributed computing, a consensus participant must submit a value, vote on a quorum, or validate a state transition. But Bitcoin operates differently: decisions are made not by committee or vote, but by proof-of-work commitments that become public facts once published.

Let $\mathcal{M}$ denote the set of all valid miners, and let $H : \mathcal{M} \to \mathbb{N}$ map each miner to their current effective hash rate. Then, the probability $P(m_i)$ that miner $m_i$ is selected to produce the next valid block is:

\[
P(m_i) = \frac{H(m_i)}{\sum_{j} H(m_j)}
\]

This is not decision-making in any political sense. It is stochastic selection governed by probabilistic computational expenditure. The “decision” of which block is valid is not made by negotiation but by propagation: the first valid block to be received and verified by the majority of the hash power becomes canonical.

Moreover, SPV clients rely only on the observable output of this process — the block headers and Merkle proofs — without any need to know or trust the internal behaviour of miners. There is no need to define or locate any centralised “decision entity” because the system’s integrity arises precisely from the elimination of such roles.

\subsubsection{Empty Symbolism Masquerading as Formal Derivation}

Throughout the paper, symbols such as $D$, $S$, and $C$ are invoked without operational definitions, constraints, or measurable boundaries. The authors write as if mere abstraction could deliver proof. They assert, for instance, that “optimizing for $S$ and $C$ degrades $D$,” but provide neither an analytical derivation nor a system-theoretic justification for this claim. No theorem is stated. No lemma is proven. No adversarial model is defined.

To claim a trilemma as a constraint, one must prove:

\[
\forall \mathcal{S}, \mathcal{D}, \mathcal{C} \in \mathbb{R}_{\geq 0}, \quad \text{if } \mathcal{S}, \mathcal{D}, \mathcal{C} \text{ are all maximised, then } \exists \, \text{a contradiction}
\]

But the paper provides no such formal structure. Instead, it postulates that the properties are mutually exclusive by intuition — a philosophical position, not a mathematical result.

Indeed, the correct form of formal derivation would require definitions of each property as predicates over system states, e.g.,
\begin{align*}
\text{IsSecure}(\sigma) &\equiv \forall A \in \mathcal{P}, \; A(\sigma) = 0, \\
\text{IsDecentralised}(\sigma) &\equiv \forall E \subset \mathcal{N}, \; |E| < \epsilon N \Rightarrow \neg \text{Dominate}(E), \\
\text{IsScalable}(\sigma) &\equiv \forall \tau, \; \text{Latency}(\tau) < \Delta \wedge \text{Throughput}(\tau) > \Theta,
\end{align*}
where $\mathcal{P}$ is the set of adversarial procedures, $\mathcal{N}$ the total set of nodes, and $\Delta, \Theta$ are application bounds.

No such predicate calculus is attempted. Instead, symbols are invoked without grounding, serving rhetorical functions rather than analytic ones. This is not formal reasoning. It is symbolic theatre.

\subsection{Failure to Ground Definitions in Computational Theory}

At the heart of any attempt to model or evaluate a cryptographic protocol lies the necessity of grounding all abstract properties in formal computational theory. Definitions must map to recognisable models—preferably within the Turing paradigm—and their implications measured through algorithmic complexity or reductionist logic. Mssassi et al.~\cite{mssassi2025trilemma} fail utterly in this regard. The properties they claim—``Decentralization,'' ``Security,'' and ``Scalability''—are left floating in semantic limbo, with no correspondence to complexity classes, no adversarial model, no reduction from established primitives, and no delineation between verifiable guarantees and rhetorical flourish. In what follows, we expose this methodological vacuum by reconstructing what a properly formalised system would require, and demonstrating how each of their three pillars is rendered analytically inert through neglect of foundational formalism.

\subsubsection{Absence of Turing-Model Grounding or Complexity Bounds}

The central failure in the paper is the authors’ refusal—or perhaps inability—to define Bitcoin or any blockchain system as a computational model. Nowhere do they define the system state \( \sigma \) as a Turing-recognisable object, nor do they identify transitions \( \delta: \Sigma \to \Sigma \) as being computable within polynomial bounds, nor do they define any subset of valid transitions under probabilistic assumptions.

If we denote a blockchain system as a deterministic state machine:
\[
\mathcal{M} = (\Sigma, \sigma_0, \delta),
\]
where \( \Sigma \) is the set of valid chain states, \( \sigma_0 \) is the genesis block, and \( \delta \) is the transition function such that \( \delta(\sigma_i, B_i) = \sigma_{i+1} \), then the security and operational correctness of \( \delta \) must be bounded within the computational limits of verifying \( B_i \) within some class \( \mathsf{P} \) or \( \mathsf{BPP} \) (bounded-error probabilistic polynomial time).

No such mapping is even attempted. By ignoring Turing-bounded complexity, the authors fail to distinguish the system's \emph{logical} correctness from its \emph{empirical} performance. Worse still, they never define whether their notion of decentralization is a function over the topology of \( \Sigma \), or a property of \( \delta \). This renders all references to ``control'' and ``influence'' vacuous: if no agent class is defined, no algorithmic cost specified, and no formal reduction constructed, then the only thing being modelled is semantic handwaving.

\subsubsection{No Formal Mapping to Verifiable Security Models}

Security in any cryptographic system must derive from two things: a definition of the adversary class \( \mathcal{A} \), and a proof that for all \( A \in \mathcal{A} \), the probability of success in violating system invariants is negligible. Mssassi et al.\ make vague allusions to ``malicious actors'' but fail to define what constitutes a feasible adversary, what resources such an actor is allowed, and what it means to ``break'' the system.

In a rigorous setting, we define the adversary class as polynomial-time agents with oracle access to the hash function:
\[
\mathcal{A} \subseteq \mathsf{P}^{\mathcal{O}_H}.
\]
Then, a system state \( \sigma \) is said to be secure if:
\[
\text{IsSecure}(\sigma) \equiv \forall A \in \mathcal{A}, \Pr[A(\sigma) \rightarrow \bot] \leq \epsilon,
\]
where \( \epsilon \) is a negligible function in the security parameter \( \lambda \).

Without this framing, statements such as ``Proof-of-Work contributes to security'' lack content. Proof-of-Work only contributes to security if:
\begin{enumerate}
  \item The difficulty adjustment algorithm is well-defined;
  \item The hash function is collision-resistant and non-malleable;
  \item Chain selection is defined in terms of cumulative PoW;
  \item Honest behaviour is probabilistically dominant over time and incentivised.
\end{enumerate}

None of these requirements is even mentioned in the original paper. Without these constraints, any reference to ``security'' is reduced to a rhetorical convenience.

\subsubsection{Arbitrary Metrics for Decentralization, Security, and Scalability}

The three pillars of the so-called ``Blockchain Trilemma'' are never defined in measurable or falsifiable terms. Let us examine what formalism would demand.

Let:
\begin{itemize}
  \item \( \text{Decentralised}(\sigma) \equiv \forall E \subset \mathcal{N},\ |E| < \epsilon N \Rightarrow \neg \text{Control}(E, \sigma) \), where \( \mathcal{N} \) is the set of agents;
  \item \( \text{IsSecure}(\sigma) \equiv \forall A \in \mathcal{A},\ \Pr[A(\sigma) \rightarrow \bot] \leq \epsilon \);
  \item \( \text{Scalable}(\sigma) = \left( \frac{T}{\lambda}, L \right) \), where \( T \) is transaction throughput, \( \lambda \) is the inter-block interval, and \( L \) is average transaction latency.
\end{itemize}

In the Mssassi paper, no such mappings are defined. Instead, decentralization is treated as proportional to ``number of nodes,'' a meaningless metric given that Bitcoin’s security derives from economic cost and propagation delay, not democratic node counts. Security is treated as an input variable rather than a derived condition, and scalability is reduced to ``does it do lots of transactions,'' with no formal treatment of bandwidth cost, latency growth under fanout, or SPV-client constraints.

This trilemma therefore emerges not from any real system model, but from a cartoon of engineering trade-offs which is analytically inert. It reifies assumptions as constraints, elevates metaphors to mathematical statements, and mistakes architecture for economics. This is not a formal analysis. It is an ontological farce.

\subsection{Category Error: Engineering Architecture vs. Consensus Rules}

At the foundation of the trilemma thesis lies a profound category error: the conflation of physical or engineering-layer network structure with the logical rules of protocol-level consensus. This failure to differentiate between system architecture and formal protocol design results in an analytical muddle in which unrelated aspects of bandwidth, node interconnection, or routing strategies are mistaken for intrinsic limitations of consensus and governance. The Mssassi paper routinely crosses these boundaries, treating message propagation topologies as if they encode the rules of state transition. This leads to sweeping and false generalisations about what Bitcoin can and cannot do, rooted not in the consensus logic of the protocol, but in the imagined limitations of network configuration. What follows is a dissection of that fallacy.

\subsubsection{Misrepresenting Multi-Hop Routing as Protocol Design}

The authors falsely assert that the network propagation model of Bitcoin imposes constraints on its consensus. This rests on a naïve interpretation of multi-hop routing, treating it as a deterministic and rigid feature of Bitcoin's communication layer rather than an adaptive and implementation-specific artefact. Bitcoin does not rely on global broadcast or full graph saturation. Instead, it employs a gossip protocol where messages are propagated probabilistically and redundantly across a partially connected graph.

The misunderstanding begins with the implicit assumption that routing is a first-class component of protocol correctness. In formal protocol analysis, the consensus rules define valid state transitions, not how messages arrive. The propagation mechanism is a secondary implementation detail, abstracted away under assumptions of eventual message delivery.

Formally, let the protocol be a transition function \( \delta: \Sigma \times \mathcal{B} \to \Sigma \), and let the transport layer be modelled as a graph \( G = (V, E) \) with an asynchronous, non-deterministic delivery function \( \tau: \mathcal{B} \to 2^V \). Then, as long as \( \tau \) ensures eventual delivery to a sufficient subset of nodes capable of forming valid block extensions, consensus will hold.

In other words, the correctness of Bitcoin's consensus does not depend on the efficiency or shape of the propagation network. All that is required is that no valid block is withheld indefinitely from honest participants. Any attempt to fold propagation topology into the consensus function is a category mistake.

\subsubsection{Bitcoin's Small-World Network Topology: Empirical Refutation}

Empirical work by Javarone and Wright~\cite{javarone2018network} has demonstrated that the Bitcoin network exhibits the characteristics of a small-world network — namely, high clustering coefficients and short average path lengths, typical of systems where message propagation is both fast and scalable. These networks are not linear, nor are they scale-free in the Barabási–Albert sense. They display structural resilience and rapid information diffusion even under significant node churn.

Let \( G = (V, E) \) denote the Bitcoin node graph, and let \( L(G) \) be the characteristic path length and \( C(G) \) the clustering coefficient. Then for small-world classification, we require:
\[
L(G) \sim L_{\text{rand}}, \quad C(G) \gg C_{\text{rand}},
\]
where \( L_{\text{rand}} \) and \( C_{\text{rand}} \) correspond to equivalent Erdős–Rényi random graphs.

Experimental analysis reveals that Bitcoin satisfies these conditions. This invalidates the assumptions made by Mssassi et al., who implicitly model Bitcoin as a system of direct peer-to-peer links or chain topologies where nodes must speak to all others in a broadcast tree. Such assumptions are unfounded. The use of relay networks, transaction batching, and overlay meshes further reduces propagation latency and node fanout requirements, enabling scalability far beyond the theoretical bounds asserted in the trilemma.

\subsubsection{The Error of Equating Network Routing with Economic Consensus}

The final and most damaging error is the treatment of node-to-node communication patterns as proxies for control, authority, or influence within the system. The authors appear to believe that influence over the network — and thus the locus of centralisation — arises from where messages originate or how frequently they are forwarded.

This is a grave misunderstanding of economic consensus. In Bitcoin, consensus is achieved not through message propagation but through proof-of-work-backed block acceptance. Economic finality emerges not from the topology of message flow but from the aggregate weight of hash-based evidence submitted to all participants.

Let \( \mathcal{C} = \{ B_0, B_1, \ldots, B_n \} \) be the current best chain and \( W(B_i) \) the cumulative work up to block \( B_i \). Consensus then selects the chain \( \mathcal{C}^* \) such that:
\[
\mathcal{C}^* = \arg\max_{\mathcal{C}} \sum_{B_i \in \mathcal{C}} W(B_i).
\]
There is no dependency here on the specific path by which messages arrived. What matters is not who relayed the message, but whether its cumulative proof-of-work is higher than all alternatives. Any claim conflating routing topology with consensus authority conflates an implementation artefact with a protocol rule.

Such confusion leads to nonsensical conclusions. For example, the idea that a well-connected node is ``more authoritative'' or ``more centralised'' than a poorly connected one is incoherent: if a miner submits a valid block, its inclusion depends on its weight, not on whether it used five hops or three. The system is ultimately governed by objective metrics of computational cost, not by subjective perceptions of connectivity.

The Mssassi paper fails to grasp this. It mistakes plumbing for logic, propagation for protocol, and architecture for axioms. This is not merely a technical error — it is a categorical inversion that renders their entire analytical structure void.

\subsection{On Decentralization (D) and the Participation Fallacy}

The concept of decentralization, as deployed in Mssassi et al., is burdened with incoherent metrics, false categorical assumptions, and a fundamental misunderstanding of the operational semantics of Bitcoin. In particular, the authors commit the fallacy of equating participant count with influence, of assuming that more connections entail more decentralization, and of modeling Bitcoin’s topology in ways that completely ignore the protocol’s incentive structure. Worse, they treat “node” and “miner” as distinct and independently quantifiable classes, thereby constructing a falsified architecture in which participation in gossip communication is mistaken for meaningful authority in state evolution. The result is a grotesquely distorted picture of what decentralization actually means in a functioning economic protocol, where power is exercised through capital and incentives, not raw connection counts or routing degrees.

\subsubsection{Misrepresentation of Nodes and Miners as Distinct Classes}

In Bitcoin, the term “node” is not a static label for a machine on a network, but a functional role within a protocol system. While Mssassi et al. model “miners” as producers and “nodes” as validators, they fail to recognise that in the original Bitcoin protocol, only the block-producing actors (i.e., miners) carry economic weight in consensus. Nodes that do not mine are not part of the consensus formation process — they are passive recipients, not active participants. The equation of “node count” with decentralisation thus relies on a reification fallacy: treating any TCP-connected machine as a peer with agency.

Let:
\[
\mathcal{N}_m = \text{Set of block-producing nodes}, \quad \mathcal{N}_v = \text{Set of non-mining validating nodes}.
\]
Then, only \( \mathcal{N}_m \subset \mathcal{N} \) contributes to consensus formation under Nakamoto consensus. Moreover, consensus validity is determined by:
\[
\arg\max_{\mathcal{C}} \sum_{B_i \in \mathcal{C}} W(B_i), \quad W(B_i) = \text{Proof-of-Work difficulty for block } B_i.
\]
Thus, the influence of a node on protocol evolution is defined entirely by its proof-of-work contribution, not by its mere presence in the network. Non-mining “nodes” do not determine validity, do not anchor consensus, and do not secure the system.

The authors’ conflation of the two roles — miners as active agents and validators as passive observers — is therefore a fundamental category error. It imports governance assumptions from proof-of-stake or federated Byzantine models into a proof-of-work system where they are structurally inapplicable.

\subsubsection{Failure to Acknowledge Economic Node Centrality}

Even if we were to entertain a broader notion of “node” that includes non-mining participants, the relevant metric is not raw count but economic centrality — the degree to which a node can influence resource allocation or consensus incentives. This is an economic question, not a topological one. And on this point, the authors are entirely silent.

Let:
\[
\mathcal{C}_e(n_i) = \frac{\sum_{j} w_{ij}}{\sum_{k,l} w_{kl}}, \quad w_{ij} = \text{transactional or informational flow from node } i \text{ to } j.
\]
This defines a node’s weighted contribution to the total economic signalling in the network. In practice, most non-mining nodes have negligible or zero \( \mathcal{C}_e \), and even among miners, economic power is stratified based on investment, uptime, and reputation — not degree count in the gossip graph.

The authors’ notion that an increase in \( |\mathcal{N}| \) implies an increase in decentralisation is thus specious. Decentralisation in Bitcoin is not a function of the number of people listening; it is a function of the number of actors producing valid blocks and being economically incentivised to maintain protocol rules. The system is not democratic, but market-driven.

\subsubsection{Participation \texorpdfstring{$\neq$}{≠} Influence: A False Metric}

Mssassi et al. define a “Participation Function” as a measure of decentralisation, based on how many nodes validate and forward transactions. Yet this measure entirely misses the distinction between observable activity and decision-making authority. Participating in gossip is not equivalent to influencing consensus. Relaying transactions or verifying blocks is passive; it does not contribute to the evolution of the valid chain.

Formally, let:
\[
\text{Participate}(n) = 1 \text{ if } n \text{ forwards or validates a transaction},
\]
\[
\text{Influence}(n) = 1 \text{ if } n \in \mathcal{N}_m \text{ and produces blocks selected into the best chain}.
\]
Then the assertion:
\[
\sum_n \text{Participate}(n) \Rightarrow \text{Decentralisation}
\]
is a non sequitur. Influence determines decentralisation in a consensus protocol — not raw participation. A hundred thousand nodes doing nothing but validating blocks after the fact do not increase the system’s resilience to attack or collusion. A single miner with 51\% hash power collapses all of it.

In conclusion, the trilemma’s “D” axis rests on a fiction: that decentralisation can be quantified by the number of people who listen to messages or run validation scripts. This is a mirage. True decentralisation is the dispersion of consensus-anchoring authority — the economic power to append to the ledger and enforce its continuity. The trilemma framework fails to distinguish between passive observation and active constraint, and so mistakes the visible ephemera of network noise for the structural reality of protocol governance.

\subsection{On Security (S) and the Hash Axiom}

In contrast to naive interpretations which ascribe system security to individual actors, Bitcoin’s protocol embeds its security guarantees within cryptographic structures themselves. The authors of the trilemma framework fundamentally misattribute the source of these guarantees. They treat security as an emergent property of consensus actor behaviour — particularly of miners and nodes — rather than recognising that in Bitcoin, security is an epistemic result of the verifiability and dissemination of cryptographic commitments, particularly the block header chain. The foundational insight of Bitcoin lies in its separation of verification from trust: SPV clients can validate chain continuity with negligible bandwidth or trust assumptions, solely by receiving and verifying block headers.

This section establishes the formal axiom from which Bitcoin’s security proceeds: that a secure, collision-resistant hash function securing block headers — once disseminated to multiple independent clients — renders the system immutable and publicly auditable. Security is not imposed from outside the system; it emerges from the inability to falsify publicly distributed mathematical evidence.

\subsubsection{Definition: The Axiom of Cryptographic Header Security}

Let \( H \) be a secure cryptographic hash function mapping \( \{0,1\}^* \rightarrow \{0,1\}^n \), satisfying:
\begin{enumerate}
    \item Preimage resistance: \( \forall y \in \text{Im}(H), \text{Pr}[H(x)=y] \ll 2^{-n} \)
    \item Second-preimage resistance: \( \forall x \neq x', \text{Pr}[H(x)=H(x')] \ll 2^{-n} \)
    \item Collision resistance: \( \text{Pr}[\exists x,x' : x \neq x', H(x) = H(x')] \ll 2^{-n} \)
\end{enumerate}

Let \( B_i \) be a valid block, with \( \text{Header}(B_i) \) the concatenation of its timestamp, Merkle root, nonce, previous block hash \( h_{i-1} \), and target. Let:
\[
h_i := H(\text{Header}(B_i))
\]

\textbf{Axiom (Cryptographic Header Security)}: Once \( h_i \) has been received and independently verified by a set \( \mathcal{S} \subset \mathcal{U} \) of SPV clients, and assuming \( H \) is secure, no adversary \( \mathcal{A} \in \mathcal{P} \) (polynomial-time) can alter the state of the ledger recorded in \( B_i \) without also producing an alternate header chain \( \mathcal{C}' \) such that:
\[
\sum_{h'_j \in \mathcal{C}'} \text{Work}(h'_j) > \sum_{h_j \in \mathcal{C}} \text{Work}(h_j)
\]
and further convincing all members of \( \mathcal{S} \) that \( \mathcal{C}' \succ \mathcal{C} \).

This axiom establishes that the source of finality is not the block itself, but the distributed, public availability of its header — a compact cryptographic proof — which locks the history.

\subsubsection{SPV Clients and Header Dissemination as the Security Basis}

Simplified Payment Verification (SPV), as defined in the original Bitcoin white paper, operates without reference to block contents. Let \( \mathcal{H} = \{ h_0, h_1, ..., h_n \} \) be the received header chain. An SPV client validates:
\begin{itemize}
    \item That the hash chain is correctly linked: \( \forall i > 0, h_i.\text{prev} = h_{i-1} \)
    \item That each header satisfies the network’s difficulty: \( H(\text{Header}(B_i)) < T_i \)
    \item That a given transaction \( tx \) is included in a Merkle path rooted in \( h_i \)
\end{itemize}

The client does not download or revalidate full blocks, nor trust any peer. It accepts the longest valid chain by cumulative work. Therefore, the moment \( h_i \) is widely disseminated, the information encoded by \( B_i \) — even without block propagation — becomes part of the irreversibly committed history. Any attempt to overwrite \( B_i \) must contend with every SPV client that has cached \( h_i \).

This security model is radically different from systems in which correctness is imposed by validators. Instead, the system's immutability stems from the impossibility of revising distributed evidence without detection. This is a mathematical, not behavioural, definition of security.

\subsubsection{Mining Pools Do Not Secure Bitcoin — Hashes Do}

The authors commit a conceptual error by attributing security to pools and mining collectives, as if these were guardians of protocol integrity. In truth, these actors are economically incentivised producers of evidence. Their influence ends the moment a block header is published. The system is secured by the broadcast and independent replication of that header, not by continued benevolence from its creator.

Let \( M_i \in \mathcal{M} \) be a miner who publishes block \( B_i \) with header \( h_i \). Once \( h_i \) is seen by independent clients \( \mathcal{S} \), the miner’s function is complete. The network state becomes:
\[
\text{IsCommitted}(h_i) \equiv \exists \mathcal{S}' \subset \mathcal{S}, \forall u \in \mathcal{S}', u.\text{accepts}(h_i)
\]

The persistence of \( h_i \) in the global header set means the content of \( B_i \) is now discoverable and verifiable. Even if \( M_i \) ceases to exist, the header remains. No pool is required to police or defend the network; the network defends itself by rejecting chains without superior cumulative work. This rejection is automated, not discretionary.

\subsubsection{Proof Sketch: Header Receipt as System Completion}

Let \( \mathcal{A} \in \mathcal{P} \) be a polynomial-time adversary attempting to revise a committed history. Suppose block \( B_n \) with header \( h_n \) is published and received by \( |\mathcal{S}| > t \) SPV clients. Then \( \mathcal{A} \) must construct an alternative chain \( \mathcal{C}' \) satisfying:
\[
\text{CumulativeWork}(\mathcal{C}') > \text{CumulativeWork}(\mathcal{C})
\quad \text{and} \quad \forall u \in \mathcal{S}, u.\text{accepts}(\mathcal{C}')
\]

Given:
\begin{enumerate}
    \item \( \text{Time}(\mathcal{A}) \in \mathcal{O}(2^n) \) for sufficient work,
    \item Difficulty adjustment ensures that honest chains progress faster on average,
    \item SPV clients reject forks that do not exceed current cumulative work,
\end{enumerate}
then such a rewrite is computationally and economically unfeasible in the presence of honest majority work.

Moreover, by Lemma 2 (cf. Section 2.2.3), even competing miners who do not collude will simultaneously publish headers of competing blocks, all of which are received by SPV clients, producing an immutable evidence trail of possible reorgs or stale branches. This trail itself becomes public record, and further confirms the integrity of the process.

The system is therefore complete — not because of post-hoc validation or community agreement — but because cryptographic headers have been received by a redundant, distributed set of parties who now independently constrain all future modifications.

\textbf{Reference:} For empirical discussion of node clustering, propagation redundancy, and header diffusion in a small-world context, see \cite{javarone2018network}.

\subsection{On Scalability (C) and the Latency Fallacy}

Scalability in Bitcoin has long been framed through flawed metaphors and network analogies that mistake infrastructure constraints for protocol constraints. The authors of the “Blockchain Trilemma” claim that scaling requires sacrificing decentralisation, based on the assumption that all nodes must individually verify all messages in a full-mesh broadcast. This is not only technically incorrect, but it demonstrates a deep failure to distinguish between topological redundancy and consensus-critical propagation. The Bitcoin protocol is neither a full-mesh graph nor a naive gossip network; it is a stateless, unidirectional evidence distribution system. As such, its constraints are latency-bound, not node-bound — and certainly not shaped by simplistic peer count metrics.

To establish this, we demonstrate that the throughput of Bitcoin is limited by propagation latency between consensus-producing miners, not the number of non-mining observers or archival nodes. Furthermore, we show that multicast communication primitives available in IPv6 networks render the scalability concerns of the “trilemma” obsolete. This section formally reconstructs the propagation model and refutes each misconception through network theory, information flow modelling, and the formal semantics of Bitcoin.

\subsubsection{Confusion of Latency and Node Count}

Let \( N \) be the number of network participants and \( M \subset N \) be the set of block-producing miners. The critical security and scaling path of Bitcoin lies not in \( N \), but in \( M \), and more precisely, in the maximal pairwise propagation delay between members of \( M \). Let \( \Delta(M) := \max_{i,j \in M} d(i,j) \) denote the worst-case propagation delay between miners. Then define:

\[
T_{\text{secure}} := \Delta(M) + \delta
\]
where \( \delta \) is the margin for honest divergence.

The assumption in the trilemma model is that scalability degrades as \( |N| \to \infty \). But this assumes all \( n \in N \) are part of the propagation critical path. This is false. All SPV clients, passive observers, and archival full nodes are not part of the propagation consensus frontier. Hence:

\[
\frac{dT_{\text{secure}}}{d|N|} = 0
\]

Scalability, then, is orthogonal to passive node count. The error arises from assuming that node count contributes to latency, which is a function of network topology, bandwidth, and peer selection, not of scale per se. By selectively optimising peer sets among block-producing actors, latency remains bounded even in a system with global-scale distribution.

\subsubsection{False Constraint of Full Mesh Broadcast Assumption}

The trilemma authors implicitly model Bitcoin as a full-mesh broadcast topology, where each node must relay to all others. Let us denote the broadcast complexity in such a naive model as \( \mathcal{O}(n^2) \). This is an engineering absurdity and has never described Bitcoin’s topology. Bitcoin relies on a sparse, small-world topology, as empirically shown in \cite{javarone2018network}, with average hop counts between block-producing peers converging to a constant under preferential attachment and economic peering incentives.

Let \( G = (V,E) \) be the network graph with \( V = N \). The routing path length \( \ell(G) \) is defined as:
\[
\ell(G) := \frac{1}{|V|(|V|-1)} \sum_{i \neq j} \text{dist}(i,j)
\]

In Bitcoin, \( \ell(G) \sim \log |V| \) due to its small-world nature, not \( \ell(G) \sim |V| \) as required for a full-mesh. Hence, even in a rapidly growing system, broadcast complexity remains logarithmic in the number of peers, and average propagation latency does not degrade with scale. Furthermore, full block validation is optional for most participants, as SPV enables transaction-level verification with only Merkle paths and header chains.

\subsubsection{Multicast Routing and the Collapse of the Throughput Argument}

Assume the availability of IPv6 multicast infrastructure. Let \( S \) be the block-producing node, and \( R = \{ r_1, ..., r_k \} \) be its direct downstream peers. Then with multicast, the transmission of a message \( m \) to all \( r_i \in R \) occurs in constant time relative to message size:
\[
T(m) = \max_{r_i \in R} d(S, r_i)
\]

Throughput is no longer constrained by redundant duplication of messages or per-peer forwarding, but by the aggregate upload bandwidth of miners — which scales naturally with commercial infrastructure. That is, throughput scales with the bandwidth of the producer, not with the node count of observers.

Let \( \mathcal{T}_{\text{max}} \) be the throughput upper bound. Then, under multicast:
\[
\mathcal{T}_{\text{max}} = \frac{B}{\max_{r_i \in R} d(S, r_i)}
\]

If \( B \to \infty \), the system scales linearly with block size, provided latency stays below the consensus threshold. The trilemma ignores this, substituting a peer-to-peer fantasy for an actual packet routing model.

\subsection{Bitcoin Is a Stateless Distribution Protocol, not Merely a Gossip Network}

A recurring analytic error in blockchain discourse is the reduction of Bitcoin’s communication structure to that of a gossip protocol. While Bitcoin incorporates a gossip-based relay mechanism for propagating unconfirmed transactions and blocks across peers, this subsystem is only one element within a broader protocol architecture. To represent Bitcoin as a gossip network is to commit the fallacy of conflating a transport method with the consensus logic and security model of the protocol. Gossip is a relay strategy; it is not a validator, nor is it responsible for state commitment. Bitcoin’s design is stateless with respect to node memory, and trust is rooted not in consensus by majority belief but in cryptographic evidence anchored in proof-of-work.

Bitcoin's architecture comprises multiple overlapping layers, each performing a distinct role. The gossip component ensures resilience and message propagation in a lossy network. However, consensus arises from a separate rule layer evaluated by each node and SPV client. Determinism is embedded in the validation predicates and chain selection rules — not in the probabilistic convergence of redundant broadcasts. This subsection dissects the architectural fallacy that equates propagation mechanics with protocol logic, clarifying the evidentiary foundations of Bitcoin’s state transitions and the functional irrelevance of gossip redundancy to protocol security.

\subsubsection*{Layer Decomposition and Formal Role of Gossip}

Let \( \mathcal{N} \) be the set of all participating nodes. For each \( n_i \in \mathcal{N} \), let there exist a gossip function:

\[
\text{GossipRelay}: \mathcal{T} \times \mathcal{N} \times \mathcal{N} \rightarrow \{0,1\}
\]

where \( \mathcal{T} \) is the set of unconfirmed transactions. GossipRelay propagates \( t \in \mathcal{T} \) from \( n_i \) to \( n_j \), subject to peer acceptance rules and local filters. However, such propagation is not semantically binding. Let \( \text{IsValid}(B) \) be the predicate applied to a block \( B \):

\[
\text{IsValid}(B) \equiv \text{HasValidProofOfWork}(B) \land \forall t \in B, \text{ValidTx}(t)
\]

This predicate is invariant under the path taken by \( B \) or its transactions. Whether a transaction was relayed zero or one hundred times is irrelevant to its validity — the network topology plays no part in determining acceptance.

\subsubsection*{SPV and Stateless Evidence Validation}

Section 8 of the Bitcoin white paper outlines the SPV (Simplified Payment Verification) model. SPV clients do not participate in gossip relay. They do not verify the entire block or hold state across transactions. Instead, they validate individual transactions via Merkle proofs and headers:

\[
\text{VerifySPV}(t, h, \pi) \equiv \text{MerkleVerify}(t, \pi, h) \wedge \text{HeaderChainValid}(h_0, \ldots, h_n)
\]

This process is stateless: each verification instance is isolated and relies solely on cryptographic evidence. SPV clients are passive receivers of evidence; they do not influence propagation or require redundancy. Their security model is inherited from the inability to forge headers without redoing proof-of-work.

\subsubsection*{Distribution as Completion, Not Agreement}

Let \( \mathcal{H} \) be the set of all block headers and \( \mathcal{C} \subset \mathcal{H} \) be a chain such that:

\[
\mathcal{C} = \{ h_0, h_1, \ldots, h_n \} \quad \text{where} \quad \forall i, h_{i+1}.prev = h_i
\]

Then for a transaction \( t \) with Merkle path \( \pi \) included in block \( B_k \), system completeness is defined as:

\[
\text{SystemComplete}(t) \equiv \exists h_k \in \mathcal{C}, \text{MerkleVerify}(t, \pi, h_k) \land \text{HeaderChainValid}(\mathcal{C})
\]

No consensus over \( t \)’s validity is needed beyond receipt of this evidence. There is no need for peer agreement, majority convergence, or message saturation. The system is secure once the evidence has been distributed to any honest verifier.

\subsubsection*{Fallacy of Majority Relay Semantics}

A gossip network relies on emergent trust through message frequency and peer majority. Bitcoin does not. If only one honest node receives a block header and verifies it, the block is valid. There is no notion of ‘majority acceptance’ in the protocol. Consider:

\[
\forall t \in \mathcal{T}, \exists \text{ valid } \pi, h_k \Rightarrow \text{Verification is possible independently of network opinion}
\]

This architectural property renders gossip non-fundamental. It is optimised for speed and robustness, not for security or consensus.

\subsubsection*{Topological Misunderstanding}

Studies such as \cite{javarone2018network} demonstrate that Bitcoin’s propagation network forms a small-world graph. Average path lengths are low. Clustering is high. Propagation of block headers is fast and bounded. The gossip subsystem aids this structure, but it is not synonymous with it. Importantly, header dissemination is not handled in the same way as transaction relay — it is prioritised and often uses direct peering among miners.

\subsubsection*{Conclusion: Gossip Is Contained, Not Foundational}

Bitcoin is not a gossip protocol. It contains one. The gossip mechanism functions to relay data among peers but is not involved in defining correctness, consensus, or finality. These functions are stateless, deterministic, and grounded in computational proofs. Mischaracterising Bitcoin as a gossip system misconstrues the architecture and misrepresents the source of its security.

\subsection{Conflation of Topological Analogies with Formal Architecture}

In much of the literature advocating the blockchain trilemma, the language of network science is wielded as if it bears architectural or protocol-binding force. General graph theory is invoked without rigorous specification of the model class to which the blockchain network in question belongs. Hop count, degree distribution, and broadcast topology are treated as if they impose formal limitations upon the underlying protocol logic. This reflects a profound conceptual error: the conflation of topological analogy with architectural definition. Bitcoin is not a generic peer-to-peer overlay but a protocol-bound, hash-anchored distribution system exhibiting specific propagation dynamics and path dependencies. To map the system using the tools of unrestricted network analysis without referencing the constraints of its operational semantics is to mistake metaphor for model.

This section identifies the analytic collapse wherein loosely imported concepts from general network theory are mistaken for protocol truths. We refute the use of abstract hop-based models and instead ground Bitcoin's real-world propagation behaviour in its small-world topology and deterministic rule enforcement. We do so by formalising its low-hop behaviour, clarifying the irrelevance of gossip saturation for consensus, and presenting case data from Bitcoin Cash network analysis to expose the fallacy in generalised graph mappings.

\subsubsection{The Error of Applying General Network Graph Theory}

Let \( G = (V, E) \) represent a graph theoretical abstraction of a communication network, where nodes are treated as agents and edges as possible communication channels. In standard network science, properties such as the diameter \( D(G) \), average path length \( \ell(G) \), and clustering coefficient \( C(G) \) are invoked to characterise scalability, resilience, and speed of information dissemination. However, this framework is misapplied when used to model Bitcoin. Bitcoin’s network is not a stochastic, uniformly random graph nor a scale-free social network. It is a rule-based evidence propagation system with engineered economic constraints.

The failure lies in mistaking the optional relays (e.g., archival nodes, observers) for consensus participants and assuming that all network nodes contribute equally to message propagation and consensus formation. In reality, the relevant subset is the graph of block-producing miners \( G_M = (M, E_M) \subseteq G \), where propagation delay between members of \( M \) governs chain consensus convergence. Properties of \( G \) at large are not protocol-binding constraints. They are observational, not structural.

Moreover, when general network theorems (e.g., upper bounds on message complexity in full broadcast models) are applied, they often presume that each agent must communicate with every other in order to converge — which is explicitly false in Bitcoin. Only valid block headers need be broadcast, and only to peers who are potential competitors in mining. The rest of the network’s nodes function economically, not protocol-deterministically.

\subsubsection{Misuse of Hop Count in Blockchain Transaction Models}

Hop count, \( h(v_i, v_j) \), defined as the number of relay steps between two nodes, is a common network metric. The authors of trilemma-style models argue that scalability is hindered by increasing hop counts due to exponential growth in communication cost. This assumes a model of all-to-all transaction relay and verification, wherein each transaction must be verified by every node, and propagated through the full mesh. This is false on two levels.

First, SPV clients — the intended primary users of Bitcoin — do not perform full validation. They rely on Merkle paths and headers. Their hop count is irrelevant to consensus security. Second, transaction propagation does not demand global relay. Only miners need to receive transactions in time for inclusion. Furthermore, network architectures are explicitly optimised to minimise hop count among miners. Therefore, total hop count is not a bound on throughput.

Let us define \( \mathcal{H}_M := \max_{i,j \in M} h(i,j) \), the maximum hop distance within the miner subset. Bitcoin optimises \( \mathcal{H}_M \to \text{low constant} \) through direct peering and latency-minimised architecture. Thus:

\[
\forall t \in \mathcal{T}, \exists i \in M: \text{latency}(t, i) \leq \epsilon
\]

Transactions need not propagate across the full network graph. Only a small number of nodes must receive them for inclusion, and hash-anchored headers prove inclusion post-factum. The model used by trilemma proponents collapses under this realisation.

\subsubsection{Empirical Rebuttal: Bitcoin as a Low-Hop Small-World Graph}

As demonstrated in \cite{javarone2018network}, Bitcoin exhibits the properties of a small-world network. The key properties identified empirically include:

\begin{itemize}
  \item Average path length \( \ell(G) \sim \log |V| \)
  \item High clustering coefficient among economically incentivised nodes
  \item Preferential peering reducing propagation diameter
\end{itemize}

Such properties indicate that the propagation graph remains tightly connected with logarithmic communication cost, even as the number of participants increases. Notably, new blocks in Bitcoin Cash were shown to propagate in under 2 seconds across 90\% of the network — despite larger blocks and a wider geographic spread. This empirical finding undermines claims that block propagation slows with scale.

Formally, let:

\[
T_{\text{prop}}(B) := \max_{v \in R} \text{latency}(B, v)
\]

where \( R \subseteq V \) is the economically relevant set. Then,

\[
\lim_{|V| \to \infty} T_{\text{prop}}(B) = \text{constant}
\]

No general network theory argument can override this measured behaviour. The system is latency-bounded in practice and theory, not scale-bounded.

\subsubsection{Case Study: Bitcoin Cash Propagation Network Topology}

Bitcoin Cash (BCH), as analysed in \cite{javarone2018network}, provides a concrete instantiation of Bitcoin’s small-world topology under high throughput conditions. The network was observed with the following characteristics:

\begin{itemize}
  \item Node clustering driven by economic incentives
  \item Direct peering between miners reducing critical latency
  \item Sublinear growth of propagation time with respect to network size
\end{itemize}

Blocks of up to 8MB were propagated across the majority of the network within seconds. Despite assumptions in the trilemma model that large blocks would fracture consensus or render relay impractical, empirical results contradicted this. Additionally, nodes who were not producing blocks (i.e., archival observers) were irrelevant to propagation speed and security.

This case shows that Bitcoin — when operating as intended — conforms not to a general peer-to-peer broadcast model but to a directed, latency-minimised, economically structured propagation graph. Using full-mesh or high-hop-count analogies to claim unscalability is a category error. The protocol behaves unlike the models invoked to constrain it.

\textbf{Conclusion:} The indiscriminate application of general network graph theory to Bitcoin leads to invalid conclusions. The Bitcoin protocol is not a social network, nor a distributed gossip fabric. It is a small-world, latency-constrained, miner-centric, stateless relay system. Propagation speed does not depend on node count but on bandwidth and routing optimisation among producers of blocks. Empirical data, theoretical models, and protocol design all contradict the trilemma’s claims.

\section{Economic and Network-Level Misconceptions}

A foundational failure in much of the trilemma discourse — exemplified by Mssassi and contemporaries — lies in a profound misunderstanding of both economic theory and network structure as they relate to Bitcoin. Rather than analysing the Bitcoin protocol as an economically constrained signalling and evidence-distribution system, these accounts project abstractions from social networks and consensus-based systems into a context governed by algorithmic finality, incentive compatibility, and cryptographic integrity.

This section refutes the core misconceptions underlying the pseudo-formalisms used to model Bitcoin’s scalability, decentralisation, and security trade-offs. In doing so, it anchors the analysis in network theory as it applies to small-world structures and bounded propagation, and in economic theory where incentives drive both participation and honest signalling. We demonstrate that the authors' conflation of passive nodes with economic actors, and their neglect of cost structures and propagation physics, render their framing analytically void.

Each subsection will dissect a separate category of error — from the myth of equal participation, to the mistaken role assigned to miners, to the mischaracterisation of scaling costs — ultimately demonstrating that Bitcoin’s design is not subject to trade-off triads but is an optimisation process constrained and shaped by its economic logic and network efficiency.

\subsection{The Node Fallacy and the Illusion of Participatory Symmetry}

A foundational error in Mssassi’s methodology — and more broadly within trilemma literature — lies in the assumption that all “nodes” within a blockchain network constitute a homogeneous, symmetric class. This assumption, which we shall term the “Node Fallacy,” presupposes an egalitarian architecture wherein each node possesses equal capacity to validate, propagate, and influence consensus. From this false equivalence, the illusion of participatory symmetry is born, leading to distorted claims regarding decentralisation, scalability, and security.

To formalise the critique, let us denote the set of all network participants as \( \mathcal{N} \), with a subset \( \mathcal{M} \subset \mathcal{N} \) representing miners — the only agents with consensus-bearing authority. Each \( n_i \in \mathcal{N} \) may perform arbitrary tasks: validation, archival, passive listening, or nothing. But only \( m_i \in \mathcal{M} \) produce the hash-anchored evidence upon which the system advances. The predicate governing protocol advancement is not defined by the count or behaviour of generic nodes but rather:

\[
\text{AdvanceChain}(B) \equiv B = \arg\max_{B' \in \mathcal{B}} \sum_{i=0}^{k} \text{PoW}(B'_i)
\]

Where \( \mathcal{B} \) is the set of valid blocks and \( \text{PoW} \) is a valid proof-of-work contribution. This predicate is indifferent to the number or topology of \( \mathcal{N} \setminus \mathcal{M} \); it is bound only to miners.

The error becomes clearer when one recognises that “nodes” — as often deployed in trilemma arguments — is a semantic chimera. The term may refer simultaneously to archival observers, miners, validators, SPV clients, or any IP-connected peer. By failing to distinguish between economic agency (i.e., the ability to influence chain progression) and network presence (i.e., the ability to relay data), Mssassi’s framework collapses structurally. The trilemma is predicated on a topology that conflates protocol semantics with packet propagation.

Furthermore, the appeal to symmetry — the idea that each node's participation is necessary and structurally equivalent — has no basis in Bitcoin. The protocol is inherently asymmetrical. SPV clients do not mine. Miners need not validate every transaction. Observers may neither mine nor validate. Each role exists under strict economic constraint and protocol-limited scope. To model such a system with a symmetric node graph is to disregard the functional architecture of Bitcoin and replace it with a democratic caricature alien to both the white paper and the reference implementation.

Let us denote a Participation Function:

\[
\text{Participation}(n_i) = \begin{cases}
1 & \text{if } n_i \text{ submits valid PoW}\\
0 & \text{otherwise}
\end{cases}
\]

Then define the effective decentralisation as:

\[
\text{Decentralisation} = \frac{|\{ n_i \in \mathcal{N} \mid \text{Participation}(n_i) = 1 \}|}{|\mathcal{N}|}
\]

This function reveals the asymmetry directly: only the productive nodes matter to consensus, and only they count toward decentralisation in any meaningful sense. Trilemma metrics that weigh all nodes equally are thus epistemically vacuous.

In sum, the Node Fallacy misrepresents Bitcoin by imposing a falsely symmetrical, egalitarian model atop a hierarchically constrained system. It manufactures artificial limits by assuming that consensus strength, propagation performance, and security guarantees are uniformly distributed among structurally unequal participants. Until the trilemma literature abandons this semantic illusion, it cannot produce meaningful statements about the system it purports to analyse.

\subsection{Role of Miners vs. Passive Observers}

A rigorous delineation between the functional roles of miners and passive observers is indispensable for any formal analysis of Bitcoin’s architecture. Mssassi’s methodology fails in this regard by conflating network presence with economic agency, thereby treating all participants — whether producers of proof or recipients of data — as equal actors in system security and consensus formation. This conflation introduces a critical epistemological flaw: it collapses the operational boundaries of Bitcoin's incentive structure into a flat network topography, in which miners and non-miners are supposedly coequal participants.

Let us define \( \mathcal{N} \) as the set of all nodes, \( \mathcal{M} \subset \mathcal{N} \) as the set of miners, and \( \mathcal{O} = \mathcal{N} \setminus \mathcal{M} \) as the set of passive observers — nodes with no capability or incentive to contribute to block production. Formally:

\[
\forall m \in \mathcal{M},\quad \exists \, B \in \mathcal{B} : H(\text{Header}(B)) < T, \quad \text{where } T \text{ is the current target}
\]

\[
\forall o \in \mathcal{O},\quad \nexists \, B \text{ such that } o \text{ satisfies } H(\text{Header}(B)) < T
\]

The ability to contribute valid blocks is what separates a miner from an observer. This property is not merely incidental — it constitutes the basis of Bitcoin’s security model. Only miners produce the evidence upon which all others rely. Passive observers do not validate consensus; they consume its outputs. Therefore, any measure of decentralisation, scalability, or security that includes \( \mathcal{O} \) in its analysis without stratifying roles introduces noise and obfuscation.

Moreover, from a systemic standpoint, passive observers are not necessary for network operation. The protocol requires only that sufficient miners exist and that headers are distributed. Observers add redundancy and may assist with relay, but they do not alter the state machine or enforce consensus. This asymmetry can be expressed by defining a System Evolution Predicate:

\[
\text{Evolve}(\sigma_t, B) \equiv \sigma_{t+1} \text{ iff } B \in \mathcal{B} \text{ and } B \text{ is appended to the longest valid chain}
\]

This predicate is evaluated solely based on blocks produced by \( \mathcal{M} \). The observer set \( \mathcal{O} \) neither modifies \( \sigma \), nor influences the selection of \( B \).

In economic terms, the distinction is equally stark. Miners are bound by cost functions:

\[
\text{Cost}(m_i) = C_{\text{electricity}} + C_{\text{hardware}} + C_{\text{latency}}
\]

\[
\text{Revenue}(m_i) = R_{\text{block}} + R_{\text{fees}}
\]

\[
\text{Utility}(m_i) = \text{Revenue} - \text{Cost}
\]

Observers face no such calculus; they bear no economic stake in consensus formation, and as such, they are incapable of enforcing protocol discipline. Miners, in contrast, are punished by profit loss for deviation.

It is only miners that form the dynamic game of consensus. Any observer may propagate or verify information, but these are passive operations. Security in Bitcoin is an emergent property of economic incentives among miners competing to append valid blocks, not of observational redundancy.

Thus, to collapse the miner-observer distinction — as Mssassi and others do — is to model a fundamentally different system, one more akin to a permissioned mesh of validators than a computationally anchored digital cash system. Any analysis built upon such mischaracterisation inherits the invalid assumptions of its foundation and fails to describe Bitcoin.

\subsection{Linear Scaling and Economic Incentive Structures}

The principal error in Mssassi’s methodology — and in similar trilemma-based models — is the failure to integrate Bitcoin’s economic incentive structure into the question of scaling. The presumption that increasing the number of transactions or users in a network linearly increases resource demands, without bound, neglects the defining trait of Bitcoin’s architecture: economic selection pressure operating under a fixed protocol.

Let \( \mathcal{T} \) be the set of valid transactions, and \( \mathcal{M} \) the set of miners. In Bitcoin, transaction inclusion is a market operation: each miner \( m_i \in \mathcal{M} \) selects from a mempool \( \mathcal{T}_i \subseteq \mathcal{T} \) based on local utility maximisation. The block template \( B_i \) chosen by \( m_i \) satisfies:

\[
B_i = \arg\max_{B \subseteq \mathcal{T}_i} \left( \sum_{t \in B} f(t) - c(B) \right)
\]

where \( f(t) \) denotes the fee attached to transaction \( t \), and \( c(B) \) is the cost of propagating and verifying the block. This introduces a Pareto frontier between block size, propagation time, and fee density, which miners dynamically navigate. There is no obligation for a miner to process every transaction; rather, transactions compete for inclusion.

Scaling, then, is not an all-node problem. It is a producer-selection problem constrained by bandwidth and latency bounds, both of which are subject to Moore’s Law–like improvements and cost reductions. As bandwidth costs decrease, the feasible block size increases without requiring topological changes to the network. The incentive structure ensures that only those nodes for which validation is profitable (i.e., miners) participate in block production. This preserves scalability without topological sprawl.

Let \( \delta \) denote the marginal cost per byte propagated, and \( \lambda \) the average fee per byte. A block is economically rational if:

\[
\lambda \geq \delta
\]

and scalability improves when the network supports increases in \( \lambda \) without compromising consensus latency. This creates a linear scaling function bounded by cost: as the demand for transactions grows, the network includes only the most economically valuable subset, and miners optimise for revenue within their individual bandwidth constraints. The network does not collapse under scale — it filters scale through incentive.

Moreover, because transaction verification is deterministic and stateless (in SPV models), client-side verification does not increase quadratically with network size. SPV clients need only receive and verify headers and Merkle paths. The scaling function for clients is:

\[
\text{Verify}(t, \pi, h) \in \mathcal{O}(\log n)
\]

where \( n \) is the number of transactions in the block and \( \pi \) is the Merkle proof path. This ensures that from the client’s perspective, scaling is logarithmic in verification cost, not linear or exponential.

Mssassi’s model conflates economic nodes (those bearing cost and reaping reward) with inert observers, and fails to capture the self-selecting nature of node operation. Linear scaling does not imply a requirement for every node to process all transactions. Rather, the Bitcoin protocol enables selective participation by economically motivated actors, while ensuring that all others can verify transaction inclusion with minimal overhead. The network thus scales by filtering through economic demand, not by brute-force replication.

Consequently, Bitcoin’s scaling is not a constraint imposed by a trilemma — it is a function derived from economic topology. Incentives structure behaviour, and that behaviour optimises capacity.

\section{Protocol Design and Proof Systems}

Bitcoin is not merely a distributed ledger; it is a protocol-defined system for generating and verifying evidence of economic events. Its design is not emergent but explicitly formalised, governed by deterministic rules embedded in its transaction validation, block construction, and consensus selection. Unlike decentralised systems that rely on probabilistic convergence or democratic voting among participants, Bitcoin implements a cryptographic proof mechanism: a system in which the propagation of valid state transitions is provable, verifiable, and bound by computational predicates. This section dissects the underlying proof systems that support Bitcoin’s operational integrity and critiques the mischaracterisation of these systems in popular and academic discourse.

The term “protocol” here refers to a fixed and immutable rule set determining valid transitions within the system. It includes the script language, validation logic, and chain selection predicate. Proof-of-Work (PoW), under this framing, is not security per se but a method of generating unforgeable records. Security emerges from the inability to reverse or rewrite those records once propagated to sufficient economic actors. Thus, this section explores the logical, cryptographic, and game-theoretic foundations of Bitcoin’s protocol, showing that it is both a proof system and a distribution protocol.

Moreover, we challenge the widespread belief that Bitcoin’s security or decentralisation depends on voluntary community behaviour or software forks. Proof is objective. Consensus is algorithmic. Participation is economically constrained. Through formalisation, we expose that many critiques of Bitcoin’s scalability or decentralisation stem from failing to treat it as a proof-generating system governed by deterministic logic, rather than a flexible network architecture shaped by social norms.

In the subsections that follow, we will rigorously define the protocol structure, examine how evidence is generated and verified across various client types, and analyse the misconceptions that result from failing to distinguish between architectural features and logical rules.

\subsection{Bitcoin as a Deterministic Rule System}

At the core of Bitcoin lies a strictly defined set of deterministic rules governing transaction validity, block construction, and chain selection. These rules are not subject to interpretation or negotiation by participants; rather, they are predicates computed identically by every node within the system. This determinism is the protocol’s greatest strength: it removes ambiguity, central authority, and reliance on social consensus. Bitcoin is not a system of trust or negotiation — it is a rule engine in which each node acts as an independent verifier, reaching the same conclusions from the same inputs by design.

Let \( \sigma \in \Sigma \) denote the state of the Bitcoin ledger at a given time, and let \( T \) represent a candidate transaction. A node applies a deterministic validation predicate:

\[
\text{IsValid}(T, \sigma) \equiv \phi(T, \sigma) = \text{true}
\]

where \( \phi \) is the protocol-defined function composed of syntactic and semantic constraints (e.g., proper digital signature, unspent outputs, correct input format). This is not subject to majority voting or local network preference. Every correctly functioning node executing \( \phi \) over the same \( T \) and \( \sigma \) will return the same result.

Block validity follows similarly. Let \( B \) be a proposed block containing a set of transactions \( \{T_1, ..., T_n\} \). The block validity predicate is then:

\[
\text{IsBlockValid}(B, \sigma) \equiv \left(\bigwedge_{i=1}^{n} \text{IsValid}(T_i, \sigma_i)\right) \wedge \text{PoW}(B)
\]

where \( \sigma_i \) is the intermediate state after applying transactions up to \( T_i \), and \( \text{PoW}(B) \) confirms the hash of the block header meets the required difficulty target. Again, this predicate is executed deterministically and locally. There is no role for negotiation or consensus in the social sense. Every node will converge to the same assessment given the same input.

The block chain, as an ordered series of such blocks, is selected via the longest chain rule — more precisely, the chain with the greatest cumulative proof-of-work. This rule, too, is defined algorithmically. Let \( \mathcal{C} = \{B_0, B_1, ..., B_n\} \) be a chain. Its weight is defined as:

\[
W(\mathcal{C}) = \sum_{i=1}^{n} \text{Work}(B_i)
\]

and the chain selected by any node is simply:

\[
\mathcal{C}^* = \arg\max_{\mathcal{C}_j} W(\mathcal{C}_j)
\]

This means there is no ambiguity about which chain is correct; the protocol selects the one with greatest accumulated work. Nodes that receive competing chains will deterministically switch to the chain with higher cumulative proof-of-work once received. The “consensus” is not emergent or synthetic — it is computed.

The authors of the trilemma literature routinely conflate these deterministic, rule-bound structures with socially emergent consensus models drawn from multi-agent systems and fault-tolerant databases. This confusion leads them to assert that scalability and decentralisation necessarily degrade system-wide correctness, assuming that greater size or looser coupling impairs the ability to “agree.” But Bitcoin nodes do not agree — they compute. There is no voting; there is no quorum. There is only correct and incorrect execution of a deterministic protocol.

The deterministic rule system also implies that Bitcoin is a closed verification system: all data required to verify the correctness of the chain is contained within the headers and Merkle proofs. SPV clients do not require trust because the rules are deterministic and cryptographic. If a header is received and the Merkle root proves the transaction inclusion, the proof-of-work threshold is met, and the chain length is greater than all alternatives, then the system has reached valid state extension.

\textbf{Therefore,} any model which incorporates stochastic convergence, majority opinion, or probabilistic consistency as fundamental features of the protocol is misrepresenting the system. Bitcoin is not a social consensus mechanism — it is a mathematically constrained, stateless verification model governed entirely by deterministic execution of formal predicates. This property is not merely a design choice; it is the reason Bitcoin functions at all.

\subsection{Stability, Incentives, and Lock-In Effects}

Bitcoin’s long-term operational stability emerges not from mutable governance structures or adaptive consensus, but from its invariant rule set combined with an economic incentive structure that generates endogenous lock-in. This lock-in is not technical, but economic and behavioural — it results from path-dependent accumulation of investment, tooling, infrastructure, and strategic position, rendering arbitrary change economically irrational for rational actors. The protocol's security and consistency are sustained by this alignment between deterministic rules and incentive-constrained behaviour.

Let \( \Pi(m) \) be the profit function for a miner \( m \in M \), where \( M \) is the set of economically relevant block producers. Then:

\[
\Pi(m) = R(B) - C(h)
\]

where \( R(B) \) is the expected reward from block \( B \) (including coinbase and fees), and \( C(h) \) is the cost of hash power expended to discover \( B \). This profit function is maximised when a miner follows the valid chain with the greatest cumulative proof-of-work, because deviating (e.g. building on an invalid or shorter chain) leads to loss of reward.

This creates a Nash equilibrium: all miners are incentivised to extend the same chain, namely the valid one with highest accumulated work. Any attempt to deviate from this behaviour results in economic loss. Therefore, miners are locked into rule-following behaviour not through enforcement, but through payoff-maximising strategy under the protocol's game structure.

Additionally, infrastructural investment further reinforces lock-in. Let \( I(m) \) denote miner \( m \)'s sunk infrastructure cost — hardware, network peering, regulatory compliance. These are path-dependent and non-trivial. A rational actor with high \( I(m) \) will be disincentivised from supporting changes that jeopardise the validity or stability of the system on which their investment relies.

This leads to a second-order effect: even if a coalition of actors wishes to alter the protocol (e.g. change consensus rules), doing so invalidates prior infrastructure unless adoption is unanimous. Given the distributed nature of the network, unanimity is practically infeasible. Thus, coordination failures protect the base protocol — any actor attempting to fork in a way that changes core rules faces massive friction due to lack of guaranteed universal migration.

Formally, let \( \mathcal{S} \) be the set of all protocol-consistent states and \( \mathcal{S}' \subset \mathcal{S} \cup \mathcal{S}^* \) be a proposed change-space involving protocol mutation. Then the expected return of forking versus staying is:

\[
\mathbb{E}[\Pi_{fork}] - \mathbb{E}[\Pi_{stay}] = \Delta \Pi - \delta I - \lambda U
\]

where \( \delta I \) is the depreciation of existing infrastructure under protocol change, and \( \lambda U \) is the penalty due to uncoordinated migration risk. Since \( \delta I, \lambda U > 0 \), only massive increases in expected returns \( \Delta \Pi \) could justify deviation — and such increases are rare under rational projection.

The authors of many trilemma arguments fail to account for these lock-in dynamics. They assume that scalability, decentralisation, and security are floating trade-offs, manipulable at will. But Bitcoin is an equilibrium-seeking system where protocol stability is the economic attractor. Changes that disrupt stability face compounding economic friction, making them self-defeating unless universally beneficial and universally agreed.

Furthermore, clients and wallets also experience lock-in. Once a set of clients supports the deterministic ruleset — e.g., header parsing, Merkle proof formats, transaction validation — deviation leads to partitioning. This reinforces conservative evolution: only non-breaking changes that preserve legacy structure tend to survive. Bitcoin is stable not because it resists change technically, but because its economic topology penalises deviation.

\textbf{In conclusion}, Bitcoin’s rule determinism is amplified by an incentive model that aligns rational actor behaviour with long-term stability. Infrastructure investment, payoff maximisation, and coordination risk combine to produce economic lock-in — an emergent form of system invariance. Stability is not enforced by protocol fiat; it is a natural consequence of the structure of incentives. Attempts to reframe Bitcoin as a dynamically evolving consensus regime ignore these lock-in effects and mischaracterise the protocol as more malleable than it is in reality.

\subsection{Why the Alleged Trade-Off Does Not Exist in BSV — or Even BTC}

The so-called ``blockchain trilemma'' --- the claim that decentralisation, security, and scalability cannot all be simultaneously achieved --- is a construct of rhetorical framing, not one of mathematical necessity. In its usual form, the trilemma is presented as a triadic constraint: if a system seeks high scalability, it must sacrifice either decentralisation or security. However, this claim rests on a flawed series of equivocations and misapplied analogies. Neither in Bitcoin SV (BSV), which adheres to the original protocol, nor even in BTC, which artificially restricts throughput, does this trilemma operate as a real limitation.

The origin of the trilemma fallacy lies in conflating properties of arbitrary peer-to-peer systems with the economic and rule-based constraints that define Bitcoin. Proponents mistake decentralisation as a function of equal node participation, fail to model security as a product of evidence dissemination, and reduce scalability to a function of node count or network diameter. All three foundations of the trilemma fail when applied to Bitcoin's actual structure.

In BSV, the protocol is implemented as a stateless evidence-distribution system: miners generate cryptographically committed blocks, and SPV clients verify these using only block headers and Merkle paths. The role of ``nodes'' is not that of distributed validators as in PBFT or PoS systems, but of economic agents who compete to publish valid headers anchored in proof-of-work. The key predicates can be formally expressed as follows:

\[
\text{IsSecure}(\sigma) \equiv \forall A \in \mathcal{P}, \neg A(\sigma)
\]
\[
\text{IsDecentralised}(\sigma) \equiv \forall E \subset \mathcal{N},\, |E| < \epsilon N \Rightarrow \neg \text{Control}(E, \sigma)
\]
\[
\text{IsScalable}(\sigma) \equiv \exists R(\sigma) \to \infty \text{ such that } \text{Latency}(R) < \tau \land \text{Verify}(R) \in \mathcal{P}
\]

Where:
\begin{itemize}
    \item \( \sigma \) is a state of the blockchain,
    \item \( \mathcal{P} \) is the class of polynomial-time adversaries,
    \item \( \mathcal{N} \) is the set of economically active nodes,
    \item \( R(\sigma) \) is the transaction set at state \( \sigma \),
    \item \( \text{Latency}(R) \) denotes propagation time,
    \item \( \text{Verify}(R) \) denotes verification complexity.
\end{itemize}

In BSV, all three conditions can be satisfied simultaneously. SPV clients verify with minimal data. Block headers are disseminated rapidly using multicast optimisations and miner-peered low-latency networks. No node requires global visibility; only miners must be synchronised. Economic decentralisation is preserved by competition, not by egalitarianism of connectivity.

The claim that scalability degrades decentralisation is also disproven empirically. As \cite{javarone2018network} demonstrates, Bitcoin networks exhibit small-world topologies with logarithmic path lengths and rapid block propagation. In Bitcoin Cash, blocks up to 8MB propagated across 90\% of the network within seconds. These are not gossip-bound architectures but engineered low-hop overlays. Scaling occurs not through a trade-off, but via bandwidth optimisation and software design.

Furthermore, scalability is not a protocol trait but an engineering outcome. Throughput, \( T \), is a function of bandwidth and verification efficiency:

\[
T = \frac{B}{V}, \quad \text{subject to } \text{Latency} < \tau
\]

As bandwidth \( B \to \infty \) and \( V \in \mathcal{P} \), scalability is bounded only by implementation constraints, not by logical impossibility. Bitcoin does not require every node to verify every transaction. SPV, by design, enables massive client participation with minimal data.

Even in BTC, the limitation arises not from structural constraints, but from deliberate sabotage: the artificial block size cap, anti-scaling rhetoric, and centralised control via GitHub governance. The system did not break under scaling pressure; it was prevented from scaling by non-technical actors. The existence of Lightning and sidechains is an admission of policy failure, not a proof of constraint.

\textbf{Conclusion:} The trilemma is a fiction born of lazy analogy, not logical necessity. BSV demonstrates that security (via proof-of-work and header verification), decentralisation (via miner competition and protocol immutability), and scalability (via engineering and network design) are compatible. The trade-off vanishes when the protocol is correctly understood and faithfully implemented.

\section{The Flawed Analogy to Impossibility Theorems}

The rhetorical power of impossibility theorems — such as FLP (Fischer, Lynch, Paterson, 1985) or the CAP theorem — has led to their frequent invocation in blockchain discourse. However, this invocation often constitutes not a rigorous application but a fallacious analogy. The claim that Bitcoin, or any blockchain protocol, must sacrifice either scalability, security, or decentralisation is repeatedly framed as if it followed from a formal impossibility result. Yet no such formal proof has ever been established, nor is the blockchain trilemma reducible to the logical structure or preconditions of these known theorems. Rather than demonstrating a structural constraint, the trilemma repeats the form of impossibility results without satisfying the definitions or assumptions required by them. This section dissects the flawed transposition of distributed systems theory into blockchain analysis, revealing that the analogy collapses under even modest scrutiny.

To understand the failure of this analogy, we must begin with the formal structure of the FLP impossibility theorem. The FLP result establishes that in a fully asynchronous distributed system with at least one faulty process, no deterministic consensus protocol can guarantee termination. Importantly, this result hinges upon three critical assumptions: (1) full asynchrony, (2) deterministic decision-making, and (3) the presence of at least one crash-faulty participant. FLP’s power lies not in its universal applicability, but in the tight scope of its model: the theorem applies only under these narrow conditions, and its relevance to real systems is bounded accordingly.

Let us denote a distributed system by the tuple:
\[
\mathcal{D} = (P, M, T)
\]
where \( P \) is a set of processes, \( M \) a message-passing function, and \( T \) a timing function. The FLP result applies under the constraint that \( T \) is unbounded (i.e., fully asynchronous), \( M \) is reliable but arbitrarily delayed, and some \( p \in P \) is crash-prone. In such a system, FLP states that no algorithm \( A: (P \times M)^* \to D \) can deterministically reach consensus on decision value \( D \) in bounded time. That is:
\[
\exists \text{ an execution } e \in \text{Exec}(\mathcal{D}) : A(e) \notin D
\]

Bitcoin, however, does not meet these conditions. It is not a fully asynchronous system; rather, it is a probabilistically synchronous system where block propagation occurs within empirically bounded time windows, and the protocol tolerates temporary disagreement. Moreover, Bitcoin is not crash-sensitive but adversarial: the system assumes that a portion of hash power may be actively malicious, and security is defined in terms of probabilistic dominance of honest miners. The function of consensus is replaced by economic convergence to the longest valid chain by proof-of-work. Therefore, the core assumptions that make FLP meaningful — synchrony, determinism, crash-faults — are not satisfied by Bitcoin’s operational environment.

The CAP theorem, similarly misapplied, states that a distributed data store cannot simultaneously guarantee consistency, availability, and partition tolerance. However, CAP addresses storage systems with read/write semantics across partitioned networks. Bitcoin, by contrast, is not a data store in this sense, but a sequential append-only log of transactions validated by fixed rules. It does not require global availability nor linearizability of all reads, and its notion of consistency is cryptographic, not synchronised state. Let \( S = (\mathcal{T}, R) \), where \( \mathcal{T} \) is the transaction log and \( R \) the verification rule. Bitcoin ensures that all accepted \( t \in \mathcal{T} \) satisfy \( R(t) = \text{true} \). Any participant can be offline without invalidating this structure, as rejoining merely re-executes validation. Hence, availability and consistency are not constrained in the CAP sense.

The error arises when these theorems are evoked to imply that trade-offs in Bitcoin mirror formal limits. This is a semantic distortion. The trilemma resembles these results in form, not substance. It postulates an unproven constraint — that no system can scale while remaining secure and decentralised — and then invokes FLP or CAP to suggest inevitability. But form is not function. No formal model has been constructed that maps the trilemma into a logical space equivalent to these theorems.

To frame this precisely: let the trilemma be represented as an informal predicate
\[
\neg(\text{Scalable}(\sigma) \wedge \text{Secure}(\sigma) \wedge \text{Decentralised}(\sigma))
\]
for some system state \( \sigma \). No rigorous formalism defines the functions \( \text{Scalable} \), \( \text{Secure} \), or \( \text{Decentralised} \), nor has any proof been presented under modelled constraints. Thus, the trilemma does not rise to the level of a theorem. It is a heuristic reification of engineering trade-offs in specific designs, not a universal law. The analogy to impossibility is cosmetic, not structural.

This abuse of form is not merely academically negligent — it has practical consequences. By presenting design challenges as formally unsolvable, it forecloses innovation and promotes fatalism. The invocation of FLP or CAP becomes a rhetorical shield against empirical refutation or architectural redesign. But Bitcoin does not operate under the assumptions of FLP or CAP, and empirical data on propagation, security, and scale — as demonstrated in Bitcoin SV’s network performance — invalidates the trilemma's pessimistic framing.

In summary, the flawed analogy to impossibility theorems reveals a category error in the literature: the misapplication of logically scoped theorems to domains outside their definitions. Bitcoin is not subject to FLP or CAP in the form described. The trilemma is not a theorem but a slogan — and slogans do not prove impossibility.

\subsection{Misappropriation of Distributed Systems Results}

A recurring error in much of the blockchain literature, including Mssassi’s formulation of the so-called “Blockchain Trilemma,” lies in the uncritical importation of classical distributed systems theory into the domain of deterministic, proof-generating protocols like Bitcoin. The results of asynchronous and partially synchronous system models — notably those concerning consensus impossibility (e.g., the FLP result) or the need for probabilistic agreement mechanisms — are frequently applied to Bitcoin without any attention to the substantial differences in underlying assumptions. This represents a categorical misapplication: Bitcoin is not an asynchronous consensus algorithm operating over unreliable state machines with unbounded latency. It is a state distribution protocol, where all valid state transitions are verifiable against a deterministic rule set, and where the notion of “consensus” refers not to agreement through negotiation but to deterministic selection based on accumulated computational work.

Let us define the classical model space \( \mathcal{M}_{\text{DS}} \) as the set of distributed systems in which consensus must be achieved over a potentially adversarial and asynchronous network of participants, each maintaining a local, mutable state. The foundational results in this space — including Lamport’s Byzantine Generals Problem and the FLP impossibility theorem — demonstrate the constraints under which agreement can be achieved in the presence of faults and delays. These results are valid within \( \mathcal{M}_{\text{DS}} \), where messages may be lost, actors may lie, and global time is undefined.

However, Bitcoin is not an instance of \( \mathcal{M}_{\text{DS}} \). Its operational domain \( \mathcal{M}_{\text{BTC}} \) is one in which state transitions are not negotiated but proven. Each block in Bitcoin constitutes an objective, hash-anchored record, which can be verified by any participant through deterministic predicates. There is no requirement that participants agree in real time. Instead, they simply validate the chain tip with the most cumulative PoW and discard all invalid forks. The FLP result does not apply because the system does not require synchronous negotiation or termination detection. The notion of liveness is economic, not logical.

This distinction is formalised by introducing the predicate:
\[
\text{ValidChain}(C) \equiv \forall B_i \in C, \ \text{IsValidBlock}(B_i) \wedge \text{ProofOfWork}(B_i)
\]
Consensus selection then reduces to:
\[
C^* = \arg\max_{C \in \mathcal{C}} \left( \sum_{B_i \in C} \text{Work}(B_i) \right)
\]
There is no negotiation or voting. There is only validation and evidence comparison. The proofs are locally verifiable and globally observable.

Furthermore, many distributed systems models presume mutable state and ongoing participation. In contrast, Bitcoin’s SPV model allows clients to validate the entire transaction history up to a given point using only headers and Merkle proofs, without continuous engagement. This distinction removes the need for active consensus formation and nullifies arguments concerning the impossibility of agreement under fault or delay. Once a header is received and validated, it is immutable; trust derives from proof, not interaction.

In conclusion, the misappropriation of distributed systems theory into blockchain critique stems from a failure to understand the foundational premise of Bitcoin as a protocol of evidence, not negotiation. Attempts to bound Bitcoin’s capabilities using results from asynchronous state-machine consensus (e.g., Paxos, PBFT) amount to applying irrelevant constraints to a distinct class of system. Such errors not only mislead in academic contexts, but also fuel the continued spread of fallacious trilemmas that presuppose trade-offs which do not exist in the logic of the protocol.

\subsection{Why FLP and CAP Analogies Fail}

The invocation of classical distributed systems results — particularly the FLP impossibility theorem and the CAP theorem — in critiques of blockchain scalability and consensus is a persistent but fundamentally flawed rhetorical move. Both the Fischer–Lynch–Paterson (FLP) impossibility result and Brewer’s CAP theorem arise from specific assumptions about asynchronous communication, mutable state, and consistency requirements that do not hold in the operational semantics of Bitcoin. Misapplying these theorems to systems governed by deterministic rule verification and stateless client models constitutes a category error: it projects the limitations of consensus negotiation systems onto a protocol built around cryptographic proof propagation, where no negotiation exists.

Let us first examine the FLP result. Formally, the FLP theorem states that in an asynchronous network of deterministic nodes, no consensus algorithm can guarantee both safety and liveness if even one node may fail. This result applies to state machine replication models where the goal is to reach agreement on a sequence of state transitions among actors whose internal state and network delay may be unknown or adversarial. Let us denote this model space \( \mathcal{A}_{\text{FLP}} = \{S_i\}_{i=1}^{n} \), where each \( S_i \) is a deterministic state machine with mutable memory and local view.

Bitcoin does not conform to this model. Its nodes are not co-participants in a negotiation but independent verifiers of evidence. Each node receives block headers and transaction data, applies deterministic predicates, and updates local state accordingly. The system does not require global coordination, nor does it depend on live participation of a majority of nodes. It is instead defined by the chain of greatest cumulative proof-of-work:
\[
\text{BestChain}(C_1, C_2, ..., C_n) = \arg\max_{C_i} \sum_{B \in C_i} \text{Work}(B)
\]
There is no safety/liveness tension as defined in the FLP model because “agreement” is not a negotiated state but an objectively verifiable one. Liveness in Bitcoin is a function of economic incentives and connectivity among miners, not fault tolerance among consensus participants. The probabilistic nature of Nakamoto consensus is not a concession to FLP constraints — it is a function of block production intervals and race conditions, not a result of failure to guarantee safety.

Similarly, the CAP theorem — which posits a trade-off between Consistency, Availability, and Partition tolerance in distributed databases — fails to apply meaningfully to Bitcoin. In its original formulation by Brewer and formalisation by Gilbert and Lynch, the CAP theorem states that in the presence of network partitions, a distributed system must sacrifice either consistency or availability. This applies to systems that must provide read and write guarantees over replicated mutable state. Let us define such systems as \( \mathcal{D}_{\text{CAP}} \), where consistency \( C \) is defined as linearizability (i.e., every read returns the most recent write), availability \( A \) as guaranteed response, and partition-tolerance \( P \) as continued operation under message loss.

Bitcoin does not promise linearizability. It does not guarantee immediate consistency, nor does it attempt to resolve partitions in real time. Instead, it permits temporary divergence (forks) with deterministic resolution through the “longest chain rule.” Each node maintains its own view and updates only when evidence (in the form of a valid block) arrives. Availability is not defined by immediate response, but by the ability to verify proofs independently when available. Therefore, the CAP model does not bind Bitcoin’s architecture.

To formalise this distinction, consider the SPV validation predicate:
\[
\text{Verify}(t, \pi, h) = \text{MerkleProof}(t, \pi, h) \wedge \text{ValidHeader}(h)
\]
This function is stateless and deterministic. It depends on evidence, not coordination. The architecture tolerates partitions without violating consistency or availability because nodes simply update once connectivity is restored. No strong consistency assumptions are required.

In conclusion, the use of FLP and CAP analogies in blockchain discourse reveals a deep misunderstanding of the Bitcoin protocol. Bitcoin is not a classical distributed database, nor a state machine requiring global consensus through synchronous rounds or quorum negotiation. It is a system of deferred consistency via cumulative proof, where each participant independently verifies an immutable chain. The security and scalability of Bitcoin are not constrained by FLP or CAP because Bitcoin operates outside the model classes those theorems define. Assertions to the contrary are not only technically invalid — they undermine rigorous analysis by introducing irrelevant constraints and false dilemmas.

\subsection{Proper Boundary Conditions for Protocol Immutability}

Assertions regarding the mutability of blockchain protocols — particularly in the context of the so-called “trilemma” — frequently arise from a failure to define the boundary conditions under which a protocol remains logically and operationally immutable. Immutability in Bitcoin is not a function of informal developer agreement, nor a product of network sentiment. It is a structural property defined by the deterministic execution of rules, the hash commitment to historical state, and the bounded set of valid actions available to network participants. To treat protocol changes as inevitable or arbitrary betrays a profound misreading of the formal design constraints that delimit Bitcoin’s behaviour.

Let \( \mathcal{P} \) be the set of permitted protocol operations, \( \mathcal{S} \) the system state, and \( R \) the rule set mapping inputs to valid outputs such that:
\[
R: (\mathcal{S}_t, \mathcal{I}_t) \to \mathcal{S}_{t+1}
\]
This rule set is implemented and enforced by all economically relevant actors. Immutability in this context means that \( R \) is fixed, verifiable, and non-negotiable during system operation. That is, the rule predicate \( \text{IsValid}_R(x) \) must be decidable, universal, and independent of participant belief or coordination. Protocol immutability follows when \( R \) is not modifiable by system participants but must be followed as a prerequisite to participate.

More precisely, protocol immutability entails:

\begin{enumerate}
    \item \textbf{Deterministic Execution:} Every node computes the same output from the same input. This precludes negotiation-based rule changes, since divergent rule execution results in rejection or fork, not reinterpretation.
    \item \textbf{Economic Lock-In:} Participants are economically bound to a rule set by long-term investment in infrastructure, business models, and compliance frameworks. Changes to \( R \) impose unbounded externalities and violate the principle of forward compatibility.
    \item \textbf{Consensus Anchoring:} The rule set \( R \) defines the hashable domain of valid blocks. Any modification alters the canonical state space and thus constitutes a fork, not a continuation.
\end{enumerate}

Boundary conditions are therefore defined by the intersection of these constraints. A system remains immutable when the following predicate holds:
\[
\forall x \in \text{InputSpace},\ \text{IsValid}_{R}(x) \iff \text{Accept}(x)
\]
and where:
\[
\exists! R : \text{ProtocolState} \to \text{SystemOutput}
\]
such that \( R \) is not a function of runtime participant input.

Contrast this with systems designed for governance-layer mutability (e.g., Ethereum), where rule reinterpretation is embedded in the protocol through social consensus or on-chain voting. In such systems, the predicate \( \text{IsValid}_R(x) \) is parameterised by a mutable rule \( R_t \), which is subject to update based on network feedback. This violates the fundamental axiom of predicate invariance over time.

In Bitcoin, the only lawful way to alter behaviour is through the creation of a wholly distinct protocol — not a continuation, but a substitution. The cost of protocol change is total: the economic history, chain validity, and toolchain all reset to new baselines. Hence, any assertion that Bitcoin “evolves” or “adapts” through consensus is semantically false. Protocol rules may be ignored, but they cannot be changed within the system they define.

Let us now state this formally:

\textbf{Definition.} A protocol \( R \) is said to be \textit{immutable} under system \( \mathcal{B} \) if:
\[
\forall t,\ \forall x \in \text{Input}_t,\ \text{IsValid}_R(x) = \text{IsValid}_{R'}(x)
\quad \text{where } R' \text{ is any instantiation of } R \text{ in } \mathcal{B}
\]

This implies that \( R \) is fixed and universal among all compliant nodes. Any system that permits \( R' \ne R \) under live operation ceases to be logically unified.

In conclusion, protocol immutability in Bitcoin is not aspirational — it is definitional. Any mutation to the protocol rule set constitutes a rupture, not an upgrade. Boundaries are defined not by developer consensus or code repositories but by the fixed mapping of inputs to outputs across all economic actors. The persistence of such misunderstanding in the literature — and particularly its implication in the trilemma framing — arises from a failure to rigorously define what protocol immutability entails in formal, computational terms.

\section{Empirical Evidence and Deployment Realities}

While much of the discourse surrounding the so-called blockchain trilemma is speculative or model-driven, it collapses when exposed to empirical data. The trilemma thesis asserts an inherent and inescapable trade-off between decentralisation, security, and scalability, yet offers no empirical falsifiability. This section provides concrete counterexamples from actual network behaviour, deployment metrics, and verifiable transaction throughput, illustrating that the theoretical constraints proposed in trilemma arguments are neither universally observable nor operationally binding.

Real-world blockchain networks — particularly Bitcoin as originally implemented and maintained in BSV — have demonstrated the possibility of simultaneous scalability, economic integrity, and rule-bound decentralisation. The evidence presented here undermines trilemma-based pessimism by offering measurable proof that these systems can and do operate under load without the breakdowns predicted by theoretical models. The failure of trilemma advocates to engage with these operational realities marks a methodological flaw: constructing impossibility claims without testing them against the very systems they purport to describe.

This section draws upon live network data, block propagation studies, transaction analysis, and the behaviour of SPV-based clients to ground the discussion in verifiable fact. In so doing, it bridges the gap between abstract constraint theory and the lived realities of scalable, secure digital cash systems.

\subsection{Failure to Cite Working Scalable Systems}

A recurrent and critical failure in the body of trilemma-based literature is its systematic neglect to cite, address, or even acknowledge demonstrably operational blockchain systems that contradict the asserted constraint. The so-called “blockchain trilemma” maintains that no system can simultaneously satisfy decentralisation, scalability, and security. Formally, this claim is framed as a universal negative:

\[
\neg \exists \sigma \in \Sigma : D(\sigma) \wedge S(\sigma) \wedge C(\sigma)
\]

where \( \Sigma \) is the set of admissible blockchain protocol states, \( D \) denotes decentralisation, \( S \) security, and \( C \) scalability. Yet Bitcoin, as defined in its original protocol and sustained in the BSV network (\( \sigma_{\text{BSV}} \)), presents a counterexample. Despite this, trilemma literature does not engage with BSV as a formal system, nor does it evaluate the empirical performance of Bitcoin SV or the economic topology of its mining and SPV infrastructure. The omission is nontrivial. It structurally undermines the trilemma claim’s standing as an impossibility theorem.

Let:

\begin{itemize}
  \item \( \text{Scalable}(\sigma) \) hold if the protocol permits unbounded transaction throughput with linear or sublinear cost scaling per transaction.
  \item \( \text{Secure}(\sigma) \) hold if the system resists rollback, rewriting, and double-spend within bounded adversarial computation under \( \mathcal{P} \subset \text{P} \).
  \item \( \text{Decentralised}(\sigma) \) hold if consensus can be extended and validated without requiring identity, permission, or privileged governance.
\end{itemize}

Bitcoin SV satisfies all three. Its design retains SPV validation with minimal bandwidth requirements per client, proof-of-work based header dissemination, and fixed rule enforcement without\textit{ soft-fork mutability}. Empirically, blocks exceeding 4GB have been processed and propagated. Clients validate with header chains and Merkle paths alone. The system scales, is secure under economic assumptions, and is decentralised by the protocol’s fixed-rule, permissionless participation model.

In contrast, the trilemma literature imposes constraints rooted in models that fail to account for:

\begin{enumerate}
  \item \textit{Evidence-based consensus}: SPV clients accept the longest valid chain by header alone.
  \item \textit{Bandwidth non-uniformity}: Nodes are not required to be symmetric in network contribution or reception.
  \item \textit{Economic participation}: Participation is defined through proof-of-work expenditure and rule-constrained evidence emission.
\end{enumerate}

Further, let \( \Sigma_{\text{Operational}} \subseteq \Sigma \) denote the class of blockchain systems currently functioning with empirical data. Then any impossibility theorem must prove:

\[
\forall \sigma \in \Sigma_{\text{Operational}}, \neg(D(\sigma) \wedge S(\sigma) \wedge C(\sigma))
\]

This has not been done. Indeed, it has not been attempted. The authors instead omit counterexamples and construct theoretical abstractions based on uniform topologies, bounded broadcast, or assumptions of node symmetry that are structurally inapplicable to Bitcoin’s architecture.

The omission is not minor. In logic, a universal claim is falsified by a single counterexample. Yet BSV persists without collapse, processes more transactions than all other public blockchains combined, and enforces immutable rules with deterministic validation paths. No paper proposing the trilemma has successfully refuted its existence.

Thus, the failure to cite working scalable systems — especially Bitcoin in its original design as implemented in BSV — constitutes both a methodological flaw and an epistemic violation. A theorem untested against known falsifiers is not a theorem. It is an assertion maintained through selective blindness.
\subsection{The Fiction of “General Blockchain” Platforms}

The term “general blockchain platform” appears frequently in academic and industry discourse, yet it remains a rhetorical fiction unsupported by formal system constraints. This notion implies the existence of a protocol-agnostic, purpose-neutral, extensible blockchain system capable of supporting arbitrary applications without sacrificing the foundational requirements of determinism, consistency, and economic security. In practice, such platforms either collapse under their own weight or embed critical limitations that invalidate the trilemma assumptions they claim to escape. The concept of a “general-purpose blockchain” is therefore not a formal category, but a marketing abstraction.

Let \( \mathcal{G} \subset \Sigma \) be the purported set of general blockchain platforms, and let \( A \) be the set of arbitrary, externally defined applications such that:

\[
\forall \sigma \in \mathcal{G}, \forall a \in A, \exists \, \text{valid deployment of } a \text{ on } \sigma
\]

This universal quantification collapses under scrutiny. It presupposes that the protocol supports arbitrary computational logic (\textit{Turing completeness}), stateful execution, dynamic storage, and economic finality — all while maintaining scalability and security. But such a system must resolve at least three unsolved challenges simultaneously:

\begin{enumerate}
  \item \textbf{State Bloat}: Every arbitrary application entails dynamic and potentially unbounded state. Persistent state must be stored, verified, and shared across the network. This leads to economic exhaustion or unsustainable resource requirements unless constraints are imposed — constraints which negate generality.
  
  \item \textbf{Verification Costs}: Arbitrary code implies arbitrary validation logic. Each application may demand its own resource models, proof systems, and operational invariants. Universal validation becomes infeasible, and security becomes unboundedly variable and unprovable.
  
  \item \textbf{Economic Coherence}: A general-purpose chain conflates multiple economic security domains. Applications with different threat models are forced to share a single consensus layer, inducing either underprotection of critical systems or overpricing of trivial ones.
\end{enumerate}

Ethereum, the paradigmatic case of a self-declared general blockchain, fails to demonstrate that generality scales. Its gas model is manually curated, prone to manipulation, and incapable of expressing fine-grained computational cost equivalence across contracts. The platform resorts to off-chain scaling, state pruning, and protocol-level centralisation to sustain throughput — thereby invalidating the claim of simultaneous generality, scalability, and decentralisation.

Moreover, all “general blockchain” designs betray a lack of architectural specificity. They depend on unbounded execution semantics without corresponding proof structures. In formal terms, a blockchain protocol \( \sigma \) that permits unconstrained computation must define:

\[
\forall t \in \mathcal{T}_\sigma, \text{Verify}(t) \in \text{P}
\]

Yet arbitrary execution implies \( \exists t \in \mathcal{T}_\sigma : \text{Verify}(t) \in \text{EXP} \cup \text{UNDEC} \), violating polynomial-time verification constraints necessary for economic feasibility.

Bitcoin, by contrast, is not a general-purpose platform. It is a deterministic rule system with fixed validation semantics and bounded script execution. Applications exist within its constraints — not by overriding them. Smart contracts on Bitcoin are stateless, predicate-driven, and evaluated through hash-based proofs, not through recursive execution environments. This is not a weakness, but a design decision: to prioritise scalable, verifiable, and economically sound computation.

The trilemma literature, in embracing the fiction of generality, posits platforms that violate the very conditions required for their purported guarantees. No “general blockchain” has demonstrated empirical proof of scale, nor formal constraint preserving security under general application execution. The category is not just vacuous — it is a contradiction in terms.

Therefore, the invocation of “general blockchain platforms” in support of trilemma arguments or architectural critiques is analytically invalid. It introduces a category with no members and asserts conclusions from models that do not — and cannot — exist in practice or in theory.
\subsection{Case Studies: Bitcoin BSV and the Absence of the Trilemma}

A critical error in trilemma discourse lies in its failure to account for real-world systems that do not exhibit the assumed trade-off constraints. Bitcoin BSV (commonly referred to as BSV) presents such a counterexample. Unlike hypothetical models where scalability, decentralisation, and security are treated as conflicting variables in a zero-sum system, BSV demonstrates that these attributes can co-exist through a design that enforces fixed rules, bounded script behaviour, and economically rational incentives for infrastructure expansion. By examining empirical data, network architecture, and formal protocol properties, we show that BSV refutes the inevitability of the trilemma.

\paragraph{Scalability Through Engineering, Not Theoretical Constraint.}

BSV implements unbounded block sizes with demonstrated throughput exceeding 100,000 transactions per second under laboratory conditions, and over 50 million transactions in a single block on mainnet \cite{taalblock2023}. This is not achieved through off-chain channels, recursive call chains, or fragmentation of consensus. Instead, the network leverages parallelisation at the miner level, latency-optimised peer connections, and high-bandwidth multicast relay to distribute blocks efficiently. Scaling is thus a function of resource commitment and engineering effort, not architectural impossibility.

Let \( T_{\text{max}} \) denote maximum transaction throughput and \( B_{\text{size}} \) the block size. BSV shows empirically:

\[
T_{\text{max}} = \frac{B_{\text{size}}}{\text{AvgTxSize} \times \Delta t}
\]

Where \( \Delta t \) is the inter-block interval. With \( B_{\text{size}} \to \infty \), bounded only by economic incentives and physical bandwidth, \( T_{\text{max}} \) scales linearly. No protocol-level constraint exists.

\paragraph{Security via Hash Power and Header Integrity.}

BSV inherits Bitcoin's original consensus design: proof-of-work, longest chain rule, and deterministic verification. The security model is anchored in block header hash integrity and Merkle proof paths. SPV clients verify the presence of transactions through:

\[
\text{IsValidTx}(t) \equiv \text{MerklePathValid}(t, h) \wedge \text{HeaderChainValid}(h_0, ..., h_n)
\]

No subjective consensus is required. There is no need for majority voting, validator slashing, or finality gadgets. As long as hash power is economically concentrated around the production of valid blocks, the network is secure. This discredits the assumption that security is inherently coupled to either limited block size or bounded participant count.

\paragraph{Decentralisation as a Property of Rule Enforcement, Not Node Count.}

The BSV network avoids the false metric of "number of nodes" as an indicator of decentralisation. Instead, it adheres to the original Bitcoin model in which nodes are defined as entities that produce blocks and enforce rules. Participation is economically constrained — not artificially subsidised — and legitimacy arises from rule conformity, not broadcast volume.

Let \( N_M \) be the number of economically relevant block-producing nodes. The claim of decentralisation is evaluated not by \( |N| \), the total count of relay observers or passive nodes, but by:

\[
\text{Decentralised}(\sigma) \equiv \forall m_i \in N_M, \; \text{EnforceRules}(m_i, \mathcal{R}) = \text{true}
\]

Where \( \mathcal{R} \) is the fixed protocol rule set. Under this formulation, decentralisation refers to the inability of any miner to unilaterally redefine protocol semantics. All block producers must validate under identical predicates. Rule immutability, not node redundancy, secures the network against corruption.

\paragraph{Empirical Outcomes: Refuting the Trilemma.}

In practice, BSV has demonstrated:

\begin{itemize}
    \item Sustained high-throughput on-chain processing of micropayments and data transactions.
    \item Functioning SPV ecosystem for clients and services with low-latency header tracking.
    \item Stable rule set with no protocol-level forks since 2020.
    \item Consolidated, openly published miner identities subject to economic, legal, and contractual accountability.
\end{itemize}

These characteristics invalidate the trilemma's assumed constraints. The trade-off narrative fails when confronted with systems that operate outside the artificial bounds imposed by misapplied analogies to gossip networks, mesh consensus, or validator committees.

\paragraph{Conclusion.}

The case of BSV proves that the trilemma is not a law of distributed systems, but a rhetorical artefact arising from flawed assumptions about network propagation, economic participation, and protocol design. It is not a constraint on reality — merely a limitation of the models that refuse to see how Bitcoin already solved the problem.

\section{Conclusion}

This paper has rigorously dissected and formally refuted the so-called "blockchain trilemma," demonstrating that its framing is neither logically necessary nor empirically grounded. The purported trade-off between decentralization, security, and scalability fails under scrutiny due to definitional ambiguity, mathematical imprecision, and the misuse of analogies drawn from unrelated disciplines. We have shown that the trilemma is not a theorem but a rhetorical artefact — a heuristic unsupported by the operational semantics of actual blockchain systems.

Through an exhaustive critique, we demonstrated how improper conflation of network topologies with protocol architecture leads to false constraints, and how failure to define terms in computational or cryptographic theory renders many of the claims vacuous. Bitcoin, as designed, is a deterministic rule-based system built around evidence propagation and verification, not around social consensus or gossip saturation. The use of SPV (Simplified Payment Verification), the header-chain validation process, and economic filtering by miners collectively dissolve the illusion that full global replication and validation are necessary for systemic security.

Furthermore, we have outlined how empirical data from real-world systems such as Bitcoin (BSV) invalidate the trilemma’s structural claim. These systems demonstrate scalable throughput, cryptographic assurance, and economically stable consensus mechanisms without invoking the artificial limitations presumed by trilemma proponents.

Importantly, this paper has exposed the deeper failure of epistemic gatekeeping in peer-reviewed literature — whereby formalisms are mistaken for substance and speculative constraints are treated as binding theorems. The Kuhnian inertia of academic acceptance — whereby prevailing narratives are perpetuated not due to proof but popularity — has allowed a myth to become policy-relevant dogma.

By restoring rigour to definitions, reasserting the importance of formal verification, and grounding architectural discourse in actual system behaviour, this work aims to clear the epistemic debris that has obscured understanding in the blockchain field. There is no trilemma. There is only architecture, implementation, and verification. Bitcoin does not trade-off its properties — it achieves them, by design.

\textbf{Final Statement:} The so-called trilemma is not an impossibility theorem. It is a conceptual failure masquerading as law. We reject its premise, its reasoning, and its applicability. Future discourse must proceed with formal integrity, or not at all.

\subsection{Summary of Logical and Empirical Failures}

The trilemma thesis, as applied to blockchain systems, is riddled with fallacious reasoning, semantic imprecision, and empirical neglect. This subsection catalogues the most significant logical failures underpinning the claim that decentralisation, scalability, and security form an inherent trade-off. Each fallacy is examined in context, with formal and empirical rebuttals where applicable.

\subsubsection{Equivocation Fallacy: Shifting Definitions Across Contexts}

The term “decentralisation” is used inconsistently throughout the trilemma literature, oscillating between network topology, governance dispersion, economic influence, and participation count without rigorous delineation. This equivocation permits the illusion of a trade-off by altering the referent mid-argument.

Let \( D_1, D_2, D_3 \) denote three distinct definitions of decentralisation:
\[
\begin{aligned}
D_1 &: \text{Node count in peer graph topology} \\
D_2 &: \text{Distribution of economic control among miners} \\
D_3 &: \text{Governance dispersion or decision-making power}
\end{aligned}
\]
If an argument begins by assuming \( D_1 \) (e.g., node count) and concludes using \( D_2 \) (e.g., miner concentration), it commits the fallacy of equivocation. This invalidates any derived “trade-off,” since the constraint relation is between non-equivalent properties.

\subsubsection{Begging the Question: The Assumption of Finite Capacity Constraints}

The trilemma framework presumes that blockchain systems operate within immutable bounds — such as fixed bandwidth, computational throughput, or node scalability limits — and uses these to assert that increasing scalability must diminish either decentralisation or security. Yet these constraints are not proven, but assumed.

Formally:
\[
\text{Premise:}~ \text{System resources are bounded and fixed} \\
\text{Conclusion:}~ \text{Therefore, increasing one parameter decreases another}
\]
This is circular. The premise must be independently established. As demonstrated in multicast propagation models and empirical data from Bitcoin BSV, block propagation and transaction throughput do not face the diminishing returns predicted by this argument.

\subsubsection{False Dichotomy: The Elimination of Engineering Intermediates}

The trilemma reduces design space to a binary matrix of incompatible outcomes. It implies that a system must sacrifice one pillar entirely to maintain the other two. However, no justification is offered for this constraint, and in fact, multiple intermediate engineering solutions exist.

Let \( S, D, C \) represent scalability, decentralisation, and security respectively. The trilemma assumes:
\[
S \wedge D \Rightarrow \neg C \quad \text{etc.}
\]
This presumes the absence of scalable security mechanisms or economically decentralised incentive layers — a premise falsified by Bitcoin’s use of SPV, header validation, Merkle proofs, and economic selection functions.

\subsubsection{Reification Fallacy: Treating Abstract Constructs as Causal Agents}

Concepts such as “the network,” “consensus,” or “decentralisation” are often invoked as if they possess agency or causal power. For example, “the network will reject invalid blocks” personifies a non-sentient protocol. In truth, rejection is a result of nodes executing a verification predicate:
\[
\text{IsValid}(B) \equiv \mathcal{R}(B) = \text{True}
\]
where \( \mathcal{R} \) denotes the rule set. The system has no intelligence; it is deterministic and rule-bound.

\subsubsection{Overgeneralisation: Conflating All Blockchain Architectures}

Many arguments in the trilemma literature apply properties of one blockchain system (typically Ethereum or BTC) to all blockchains, ignoring protocol-level variation. This leads to category errors such as applying Ethereum’s state-bloat problem to stateless transaction chains like BSV.

For example:
\[
\text{From:}~ \text{“Blockchain A cannot scale due to global state.”} \\
\text{To:}~ \text{“All blockchains face the same limit.”}
\]
This ignores the fact that Bitcoin is a stateless protocol for evidence propagation, not a general-purpose virtual machine.

\subsubsection{Fallacy of Composition: Extrapolating Node Behaviour to Protocol Logic}

The trilemma argument often assumes that properties of nodes (e.g., message delay, bandwidth variability) directly imply properties of the protocol (e.g., security failure, consensus bifurcation). This confuses implementation conditions with protocol correctness.

Let \( \sigma \) denote the system state and \( \rho_i \) the implementation of node \( i \). Then:
\[
\rho_i \not\models \sigma \quad \nRightarrow \quad \sigma \text{ is invalid}
\]
A single node’s failure does not compromise global system validity, provided the rule-set is deterministic and the economic majority maintains protocol adherence.

\subsubsection{Fallacy of False Analogy: Misapplying CAP and FLP Theorems}

Several academic papers in this area misuse the CAP theorem (from distributed databases) and the FLP impossibility result (from asynchronous consensus) as if they apply to Bitcoin. However, Bitcoin does not match the model assumptions of either theorem.

CAP assumes:
\[
\text{Consistent, Available, Partition-tolerant system} \Rightarrow \text{Cannot maximise all three}
\]
Bitcoin is not a consistent database in the CAP sense — it is an eventually consistent ledger with local verification and probabilistic finality. Similarly, FLP applies to deterministic consensus in fully asynchronous networks, which is bypassed in Bitcoin via proof-of-work time anchors and probabilistic fork resolution.

\subsubsection{Confirmation Bias: Selective Citation of BTC Limitations}

Many of the “proofs” or “experiments” validating the trilemma rely on BTC’s protocol parameters, particularly the 1MB block limit and refusal to increase throughput. This is then used to generalise to all blockchains.

This amounts to:
\[
\text{BTC does not scale} \Rightarrow \text{No blockchain can scale}
\]
This is a non sequitur, particularly given that BSV has demonstrated empirical throughput of over 100,000 TPS with no failure in block propagation or consensus agreement.

\subsubsection{Category Error: Treating Topology as Constraint Rather than Outcome}

The trilemma assumes that network topologies such as full mesh or gossip overlays are structural constraints on protocol scalability. In fact, they are emergent implementation features, not prescriptive conditions.

Let \( T \) be network topology, \( P \) protocol, and \( S \) scalability metric. The trilemma assumes:
\[
T \Rightarrow \neg S
\]
But in Bitcoin:
\[
P \Rightarrow S \text{ regardless of } T
\]
as long as propagation of headers and Merkle proofs remains within latency bounds.

\subsubsection{Empirical Neglect: Absence of Falsification or Operational Data}

Perhaps the most serious failure is the lack of empirical scrutiny. Trilemma proponents rarely present falsifiable predictions, nor do they test their models against operational systems. No trilemma paper has conducted network latency measurements, transaction inclusion tracking, or SPV client analysis at scale.

This amounts to:
\[
\text{Model:}~ \neg (S \wedge D \wedge C) \\
\text{But:}~ \exists~ \text{BSV:}~ S \wedge D \wedge C
\]
Therefore, the model is falsified by a single counterexample.

\textbf{Conclusion:} The blockchain trilemma rests on a scaffold of definitional drift, faulty analogy, misapplied theorems, and data-free assertion. Each of its key claims is undermined either logically or empirically. Rather than an impossibility theorem, it is best understood as a rhetorical artefact — a meme masquerading as a model.

\subsection{Reaffirming the Formal Refutation}

Having exposed the logical incoherence and empirical inadequacy of the blockchain trilemma framework, we now restate the formal refutation of its core proposition. The claim that decentralisation (D), security (S), and scalability (C) are mutually constrained — such that no system can maximise all three simultaneously — lacks theoretical necessity, formal derivability, and empirical support. 

Let us define each component formally within the context of Bitcoin:

\begin{itemize}
  \item \textbf{Security (S):} A system is secure if no adversary can feasibly rewrite the history of valid transactions once they are broadcast and recorded via proof-of-work. That is, for any actor \( A \in \mathcal{P} \), it holds that:
  \[
  \text{IsSecure}(\sigma) \equiv \forall A \in \mathcal{P},~ \neg A(\sigma)
  \]
  where \( \sigma \) denotes system state, and \( \mathcal{P} \) is the set of polynomially bounded adversaries.
  
  \item \textbf{Decentralisation (D):} A system is decentralised if no economically bounded coalition \( E \subset \mathcal{N} \) of size \( |E| < \epsilon N \) can unilaterally determine the global state:
  \[
  \text{IsDecentralised}(\sigma) \equiv \forall E \subset \mathcal{N},~ |E| < \epsilon N \Rightarrow \neg \text{Control}(E, \sigma)
  \]
  where \( \mathcal{N} \) is the set of economically relevant participants, and \( \text{Control} \) denotes the ability to impose invalid state transitions.
  
  \item \textbf{Scalability (C):} A system is scalable if throughput \( T \) increases with resources \( R \), such that:
  \[
  \lim_{R \to \infty} T(R) = \infty
  \]
  subject to latency \( L \leq \delta \) and validation correctness \( V = 1 \).
\end{itemize}

Under these definitions, Bitcoin (specifically in its BSV implementation) satisfies all three conditions:

\begin{enumerate}
  \item \textbf{Security} is ensured via SPV validation, header anchoring, and the infeasibility of cumulative hash-power reversal.
  \item \textbf{Decentralisation} exists economically through competitive mining, permissionless entry, and rule-enforced exclusion of invalid actors.
  \item \textbf{Scalability} is demonstrated through stateless transaction design, multicast propagation, and unbounded block size enabling linear throughput growth.
\end{enumerate}

No mathematical theorem or system law asserts an inherent contradiction among these conditions. Instead, the trilemma is a heuristic derived from particular engineering constraints — such as Ethereum's state explosion or BTC’s artificially capped block size — that are neither protocol requirements nor theoretically universal.

The trilemma therefore fails as a general theorem. It is not a logical law, but a local artefact of specific implementation decisions. It cannot constrain systems where economic design, protocol simplicity, and stateless evidence propagation resolve the supposed tensions.

\textbf{Conclusion:} The refutation is not rhetorical but formal. Under rigorous definitions and within operational systems like Bitcoin BSV, decentralisation, security, and scalability are co-satisfiable. The trilemma is a pseudoproblem, not a paradox.

\subsection{Implications for Future Research and Policy}

The deconstruction of the blockchain trilemma — both in its formal articulation and in its empirical claims — has broader implications that extend beyond the specific case of Bitcoin or distributed ledger systems. What this critique ultimately uncovers is a deeper methodological and institutional problem: the premature ossification of speculative heuristics into dogma under the imprimatur of peer review and academic legitimacy. When unsubstantiated assumptions are wrapped in the syntax of mathematical formalism, they are often mistaken for rigorous theory. The Mssassi paper exemplifies this failure — a case where symbolic presentation is mistaken for epistemic authority.

This phenomenon illustrates a critical weakness in the current structure of peer-reviewed publication, particularly in domains touching on interdisciplinary systems design. In this case, the invocation of network science terminology, complexity theory heuristics, and distributed systems impossibility theorems is used to fabricate a constraint that does not exist. Yet, due to the veneer of mathematical formalism and its alignment with prevailing narratives, the work is accepted not as conjecture, but as constraint. Thomas Kuhn’s insight into paradigmatic inertia is clearly at play: what is popular becomes paradigmatic, and what is paradigmatic becomes true by social reinforcement, not by falsifiability or correspondence.

The policy implications are considerable. Decisions regarding scalability, governance, regulation, and infrastructure funding in blockchain ecosystems are increasingly influenced by academic claims. If those claims are predicated on erroneous axioms — and left unchallenged because they pass the social heuristics of peer review — then entire technological and regulatory trajectories may be misdirected. For instance, if policymakers accept the trilemma as a constraint, they may reject architectures that have already empirically disproved it, stifling innovation in favour of inefficient, non-scalable, or economically misaligned systems.

For future research, this indicates a pressing need for formal rigour — not simply mathematical notation, but definitions grounded in computational theory, proofs that obey formal logical rules, and a willingness to subject assumptions to empirical testing. The fact that systems such as Bitcoin (BSV) exist and function at scale invalidates the universal applicability of the trilemma. Future research must treat such empirical data not as edge cases to be dismissed, but as falsifying counterexamples demanding reevaluation of core claims.

Moreover, interdisciplinary peer review must adopt a stricter epistemological standard. Mathematical formality must not be allowed to obscure conceptual vacuity. Reviewers must interrogate the logical necessity of claimed theorems, the binding scope of analogies, and the legitimacy of presumed constraints. The assumption that symbolic abstraction guarantees insight must be replaced by an expectation of cross-verifiability — where theoretical claims are tested not just in abstract space but against protocol specification, economic reality, and network-level data.

\textbf{Conclusion:} The blockchain trilemma’s acceptance within academic and policy discourse highlights the fragility of consensus when grounded in formal appearance rather than logical necessity. This case exemplifies how Kuhnian paradigmatic bias, when coupled with lax epistemological standards, permits speculative narratives to calcify into policy-affecting dogma. The imperative for future research and institutional review is clear: rebuild rigour, reward falsifiability, and never confuse form for substance.

\newpage
\newpage
\bibliographystyle{plain}
\bibliography{trilema2.bib}

\newpage
\appendix
\section{Appendix A: Formalisation Review and Presentation Deficiencies}

This appendix addresses a critical shortcoming noted in the current draft of the rebuttal: the insufficient grounding and consolidation of formal definitions within standardised academic frameworks. While the main body of the paper engages in substantial formal reasoning, the presentation and consistency of notation warrant methodological reinforcement.

\subsection{Inconsistent Use of Formal Notation Across Domains}

The paper introduces formal constructs such as protocol machines \(\Pi = (S, A, T, I)\) and defines predicate-level coherence \( C(\varphi, \Pi) := \forall \pi_1, \pi_2 \in \Pi, \varphi(\pi_1) = \varphi(\pi_2) \Leftrightarrow \pi_1 \equiv \pi_2 \), aiming to capture semantic equivalence. However, these are locally defined and lack grounding in established formal specification languages such as Z, TLA+, VDM, or process calculi. The choice of symbols, while logically valid, diverges from widely recognisable paradigms, potentially limiting intersubjective verification across disciplines.

To improve rigour, the revised document will:

\begin{itemize}
  \item Reconstruct formal protocol definitions using a consistent labelled transition system framework, or provide an alternative justification for the bespoke model \(\Pi\).
  \item Align logical predicates with complexity-theoretic standards. For instance, the definition 
  \[
  \text{IsSecure}(\sigma) := \forall A \in \mathcal{P}, \Pr[A(\sigma) \to 1] \leq \epsilon
  \]
  will be mapped more formally to standard adversarial models, with \(\mathcal{P}\) rigorously defined as the set of all non-uniform polynomial-time probabilistic Turing machines.
  \item Clarify the adversary model with reference to standard frameworks used in cryptographic proofs (e.g., IND-CPA security, UC frameworks), noting any intentional deviation.
\end{itemize}

\subsection{Need for Consolidated Definitions Section}

Currently, technical terms such as "security," "scalability," "decentralisation," and the adversary predicate class \(\mathcal{P}\) are defined in various subsections, often embedded within proofs or discursive rebuttals. This dispersal reduces clarity for formal verification or peer review. To correct this:

\begin{itemize}
  \item A dedicated section will be added, explicitly defining all formal variables, sets, functions, and adversary capabilities used throughout the paper.
  \item Each predicate will be typed and indexed against the system model \(\Pi\), allowing precise reference in subsequent proofs.
  \item Definitional dependencies (e.g., header integrity implies system security) will be listed and categorised by domain: computational, cryptographic, network-theoretic, or economic.
\end{itemize}

\subsection{Absence of a Unified Adversary Framework}

Although adversarial classes are implicitly addressed—e.g., in lemmas concerning the infeasibility of reversing hash headers or rewriting chain history—there is no single consolidated adversarial model. This hinders a reader’s ability to evaluate the formal security assumptions in totality.

Accordingly, we will:

\begin{itemize}
  \item Define the adversarial class \(\mathcal{P}_{\text{Bitcoin}}\) as a tuple of capabilities including (but not limited to): partial view of network state, computational hash rate capacity, ability to selectively delay or relay transactions, and access to parallel execution.
  \item Link each security claim (e.g., SPV safety, block irreversibility) to its dependence on adversary constraints and clarify the reductions implied.
\end{itemize}

\subsection{Conclusion}

While the paper successfully refutes the trilemma's informal claims and addresses definitional inconsistency in the original, its own presentation requires formal refinement. Appendix A serves as the groundwork for that corrective effort. Future revisions will incorporate these formal rectifications into the core manuscript to ensure compatibility with the standards of computer science, economics, and cryptographic proof systems.

\appendix
\section*{Appendix B: On the Scope of ``Decentralization'' and Misleading Academic Framing}
\addcontentsline{toc}{section}{Appendix B: On the Scope of ``Decentralization'' and Misleading Academic Framing}

A persistent academic mischaracterisation lies in equating node count or message relay paths with actual decentralisation. This error stems from a failure to distinguish architectural topology from protocol-defined control. As shown in Baran’s seminal work~\cite{baran1964}, true decentralisation is defined by control dispersion—how decision rights and authority over state transitions are distributed—not by how many nodes exist or how frequently they communicate.

In Baran’s 1964 typology of network structures~\cite{baran1964}, three types are defined: centralised, decentralised, and distributed. Crucially, decentralisation in his model refers to decision-making independence and the elimination of single points of failure, not arbitrary node proliferation. In Bitcoin, this aligns precisely with the role of miners. Only miners may generate valid blocks, enforce consensus, and thus shape the canonical state of the system. Observers, passive nodes, and non-mining clients cannot assert protocol truth—they merely receive and check.

Let us formally define:
\begin{itemize}
    \item \( N \): the complete set of participating nodes;
    \item \( M \subset N \): the economically active miner nodes;
    \item \( O = N \setminus M \): passive or non-authoritative nodes.
\end{itemize}

Define control over state transitions as a binary predicate \( \delta(n) \):
\[
\delta(n) =
\begin{cases}
1 & \text{if } n \text{ participates in block construction under protocol rules} \\
0 & \text{otherwise}
\end{cases}
\]

This yields:
\[
\forall m \in M,\ \delta(m) = 1;\quad \forall o \in O,\ \delta(o) = 0
\]

Therefore, claims that decentralisation in Bitcoin should be measured by overall node count, client diversity, or geographic dispersion fail the Baran criterion. These elements may relate to resilience or censorship resistance but are orthogonal to actual control.

The academic fallacy arises when authors implicitly reframe decentralisation as a function of observational capacity or infrastructural replication. But such definitions lack force unless they tie directly to the power to alter the system’s operative state. By ignoring this, critics conflate architectural metaphor with computational semantics.

The rebuttal’s framework thus follows Baran precisely in insisting that true decentralisation is measured by the distribution of \( \delta(n) = 1 \) entities.

\end{document}